# Lessons from a Large-Scale Assessment: Results from Free Response Pre- and Post-testing in Electricity and Magnetism

Beth Thacker, Keith West, Ganesh Chapagain, Vanelet Rusuriye and Hani Dulli, Physics Department, Texas Tech University, Lubbock TX 79409

## Abstract

As part of a large-scale assessment project at a large university, we administered weekly pre-tests and bi-weekly post-tests in the recitation sections of our introductory classes over four semesters from Spring 2010 through Fall 2011. The post-tests were administered as graded quizzes and were developed to assess problem solving, laboratory, calculational and conceptual skills that had been the focus of instruction in lab, lecture and recitation sessions. They were not comprehensive, but gave us "snapshots" of students' abilities throughout the semester. They were used in conjunction with other forms of assessment, such as conceptual inventories, to give us a broader picture of the state of our undergraduate classes, recitation sections and laboratories. The written pre- and post-tests, which required students to show their work and explain their reasoning, yielded different information on students' skills than the conceptual inventories. On almost all of the questions, the students in the inquiry-based course performed better than all of the other students in both the calculus-based and algebra-based classes. This study indicates that the inquiry-based students perform better at free-response questions that require students to show their work and explain their reasoning, even though their conceptual inventory scores are not significantly different from students in large lecture sections taught with Physics Education Research-informed (PER-informed) laboratories, recitations and lectures, as reported in a previous paper. This indicates a need for different and/or more comprehensive assessment instruments to evaluate the effectiveness of introductory materials and instructional techniques that can be compared across classes and universities.

# I. INTRODUCTION

While conceptual inventories in physics abound,[1] instructors and departments who have chosen to implement research-based and reformed instructional methods and materials often find themselves lacking instruments of sufficient breadth and depth in various cognitive skills, as well as content coverage, and sufficiently aligned with the goals of both reformed and traditional courses to serve as effective assessment instruments for their needs. The conceptual inventories, designed for research, are not meant to be either comprehensive or the sole indicator of success of a physics course. However, they are increasingly used for that purpose. While they should play a role in assessment, they are limited and also have drawbacks.[2-3] Many instructors not satisfied with conceptual inventories design or search for additional assessment instruments that go beyond conceptual inventories, are more in-line with the goals of their courses, cover a broader range of skills or that can be used to compliment or supplement conceptual inventories.

At Texas Tech University (TTU), we were studying an attempt at introducing materials and instructional methods informed by physics education research (PER-informed materials[4]) into the laboratories and recitation sections. We had an interesting situation in that we were in a department in which most of the instruction had been traditional and a significant number of faculty were hesitant, ambivalent or even resistant to the introduction of such reforms. However, in 2008, a small subset of instructors with an interest in reform and the support of the Department Chair at the time, introduced PER-informed materials into the laboratories and recitation sections. As there were still faculty hesitant or resistant to reforms, the lecture instruction was left unchanged and we had a situation where changes were implemented in the laboratories and recitation sections only. In Fall 2009, we were awarded a National Institutes of Health (NIH) Challenge grant[5] to support the large-scale assessment of the changes being made in the laboratories and recitation sections and we were in a position of identifying appropriate assessment instruments.

As there are very few, if any[6], valid and reliable comprehensive assessment instruments, research-based or not, in general use designed explicitly for the university level introductory physics courses, we chose, because of their validity, reliability and common use to administer a number of conceptual inventories (including the Force Concept Inventory (FCI),[7] the Brief Electricity and Magnetism Assessment (BEMA),[8] the Mechanics Baseline Test (MBT)[9] and the Conceptual Survey of Electricity and Magnetism (CSEM)[10]). We also used assessments such as Scientific Attitude and Scientific Reasoning Inventories[11] and TA Evaluation Inventories[12] for other purposes. However, we were very much aware that the conceptual inventories were not comprehensive and that they were not the best instruments to use to assess many cognitive, calculational, laboratory and other skills. We opted to administer locally written free-response pre- and post- tests in the recitations, as a means to better assess and illuminate students' calculational skills, their ability to express scientific thought in writing, and their critical thinking skills. We believed that the combination of the conceptual inventories and the free-response pre- and post- tests would give us a broader and deeper understanding of our students conceptual understanding, thought processes, problem solving abilities and laboratory skills. Also, we wanted to compare the results of free-response testing to those of the conceptual inventories and compare the results of free-response testing across classes.

In addition to the changes being made in the large lecture classes, we wished to assess a laboratory-based, inquiry-based course[15-18] that was developed with National Science foundation (NSF) funding about 10 years ago and has been taught as a special section of the algebra-based course every semester since then. It was developed explicitly for health science majors, taking their needs, learning styles, backgrounds and motivations into account. It is taught without a text in a Workshop-Physics style[19] environment and is an inquiry-based course in the manner of Physics by Inquiry,[20] developed by the Physics Education Group at the University of Washington, but at the algebra-based level. The materials were developed by modifying and adapting parts of existing materials designed for other populations and integrating them with new units in our own format, creating a course aimed specifically at health science majors.

The goals of the inquiry-based course are different from the goals of the traditional course, including a greater focus on thinking like a scientist and critical thinking skills. We expected that those skills would be more likely to show up on a different kind of assessment than conceptual inventories. The inquiry-based students, for example, are constantly required to show their work and explain their reasoning and are graded on their solution process, not just their final answer, as is often the case in the large lecture sections. The format of the course, the sequencing and also the testing is different from the standard sequence. We decided, however, in this series of pre- and post-testing, to use the pre- and post- tests that had been designed for the recitations in the large calculus-based and algebra-based courses. The pre- and post-tests were not an exact match to the inquiry-based course in either content or other skills tested, as they were to the large lecture section courses. However, we still predicted that the inquiry-based students would perform better on the free-response pre- and post-tests and that we would see differences from the results of the conceptual inventories.

We report on the results of free response pre- and post-testing in the recitation sessions of the second semester of our large lecture courses in our introductory sequences, both algebra-based and calculus-based. We present the results of select, representative, free-response pre- and post-tests administered across all of the classes in the introductory sequence that covers electricity and magnetism and optics. In addition to the results from the large lecture classes, we present results from the small, laboratory-based, inquiry-based course.[15-18]

We discuss, in Section II, the department and student populations, in Section III, the development and administration of the assessment instruments, in Section IV, the analysis methods, in Section V, the results and, in Section VI, we have a Discussion and Conclusion.

## II. THE DEPARTMENT AND STUDENT POPULATIONS

Texas Tech University (TTU), is a large university of about 32,000 students, with 26,000 of them undergraduates. The physics department has 20 tenured/tenure-track faculty and teaches about 2,600 students in the introductory physics courses each year. This includes the calculus-based and algebra-based introductory physics classes. About 1800 of these students are in the calculus-based course and 800 in the algebra-based course. The introductory courses are usually taught by faculty, but may be taught by postdoctoral researchers, visiting faculty, or even graduate students on occasion.

### A. The laboratories and recitation sections

Prior to Spring 2008, the introductory courses consisted of three hours of lecture and two hours of laboratory work each week. The labs were taught by teaching assistants (TAs). They were very traditional, "cookbook," in format and pedagogy and had not undergone significant change in many decades. The students would work through the labs and turn in a formal lab write-up to the TA. There was no recitation.

After a transitional semester in Spring 2008, PER-informed laboratories and a one-hour weekly recitation section were introduced into the algebra-based courses in Fall 2008. At first, the labs and the recitations were held in three-hour blocks, with two hours of lab and one hour of recitation. This was done mostly to help with scheduling problems as the recitations were added in. In the first course of the algebra-based sequence (ABI), the Module I of the Real Time Physics labs[21] was used exclusively. In the second course in the sequence (ABII), some Real Time Physics labs (from Modules 3 and 4) were used and some locally written PER-informed labs were used. By Fall 2010, the ABII labs were almost completely locally written PER-informed labs. They did not require a formal lab write-up, but included laboratory homework. There were also bi-weekly quizzes in the recitation sections that included material from lecture, lab and/or recitations.

Beginning in Spring 2009, a one-hour recitation was also implemented in the calculus-based courses, which we will refer to as Calculus-based I and II (CBI and CBII). The CBI labs used some of the Real Time Physics laboratories and some of the traditional laboratories. The CBII course remained traditional labs. The labs in CBI remained partially Real Time Physics and partially traditional and the labs in CBII remained traditional until Fall 2010. Starting in Fall 2010, up to the present, the labs in the second course in each sequence (ABII and CBII) were almost completely locally written PER-informed labs. The labs in the first course in the calculus-based sequence used Real Time Physics labs exclusively in Fall 2010. After Fall 2010 to the present, labs developed at the University of Illinois[22] were used.

With the introduction of PER-informed labs and recitation sections, the TAs were trained in different pedagogies than were used in the traditional labs. Most of the PER-informed labs were designed in a format that required interactive-engagement (IE) during the lab, to help guide the students.  The labs did not require a formal write-up, but included laboratory homework. There were also bi-weekly quizzes in all of the recitation sections that included material from lecture, lab and/or recitation.

1. **PER-informed labs**

    The locally written PER-informed labs consisted of five parts: Objectives, Overview, Explorations, Investigations and Summary. The Objectives listed the concepts and skills the students should understand and be able to demonstrate after completing the lab. The Overview was a short summary of the purpose of the lab. The Explorations were qualitative measurements or, sometimes, qualitative problems or thought experiments designed to focus on concepts the students may still have difficulty with, even after instruction. They might, for example, focus on drawing magnetic field lines or observing the direction compasses point near a current-carrying wire with and without current flowing through the wire.  They allowed students to experimentally observe concepts they had studied and repeat skills and review concepts that would be needed in the Investigation part of the lab.  The Investigation part of the lab consisted of quantitative measurements and observations, taking data, graphing, analyzing and interpreting it. Students would, for example, measure the magnitude of the magnetic field at different distances from a current-carrying wire and plot the data. It is the section that is more like a traditional lab.  In the Summary, the students were asked to focus on a particular part of the lab and summarize it. There was a lab homework to be completed and turned in at the next lab, but no formal lab report. A sample lab is included in Appendix I.

2. **Recitation sections**

    The recitation sections were about 50 minutes long and were usually group problem solving sessions monitored by the TA. The problems were chosen by the lab coordinator(s) and were often chosen from or modified versions of published PER-informed problems, such as problems from Tasks Inspired by Physics Education Research (TIPER)[23], Ranking Task Exercises in Physics[24], books by Arnold Arons[25-26] and other sources. Sometimes the problems were textbook problems or modified textbook problems. The problems were chosen to be on content that had already been covered by all of the instructors teaching the course. The problems were chosen to cover concepts or skills that students often struggle with, even after instruction.

The students would work through the problems in groups, working on whiteboards, with the TA circulating, asking students questions or answering questions from students. After students had had a significant amount of time to work on the problem, the TAs checked on students' understanding in different ways. Some TAs worked with groups individually, checking on their results both as they worked and as they finished, asking them to explain their results and asking further questions, as needed. Others called the class together and had groups present at the board and had a class discussion about the problems.

If there was time after the problem(s) for that week had been finished, the TAs entertained questions on homework or other questions students might have. The bi-weekly quizzes were also administered during the recitation sessions.

3. **TA training**

The TAs were trained and directed by the lab coordinator(s). They were taught to guide the students through questioning, not "telling" answers, but helping students to think through the questions themselves. They were taught how to help each group and to make sure everyone in the group contributed and was responsible for their own understanding. They were also taught how to guide whole class sessions, having groups or students present at the board and then lead class discussions. The teaching methods were modeled in their own TA training by the lab coordinator(s). The use of these IE methods was expected of them both in the recitations and in the Exploration parts of the laboratories.

**B. The faculty**

The majority of physics faculty members teach traditionally in a lecture-style format. Very few use PER-informed pedagogy or IE techniques. They focus primarily on the lecture and leave the recitations and laboratories to the TAs and lab coordinator(s). Although the labs and recitations were part of the course, the labs and recitations together were sometimes allotted as little as 10% of the grade. The lower allotments of the percentage of the grade for lab and recitation together were primarily in the calculus-based classes. However, some of the instructors allotted 20%-30% of the grade in those classes for lab and recitation (together). In the algebra-based classes, a higher percentage of the grade was allotted to the labs and recitation sections, with 20% the most common, although they ranged from 15% - 25%.

A few instructors interacted with the TAs in lab and recitations, contributing to the training of the TAs, the choice of materials and content to be covered in recitations and the pedagogy to be used in lab and recitation. Most of the instructors who actively participated in the TA training, were instructors who used PER-informed materials and instructional techniques in the lecture.

The instructors labeled by PER in this paper used PER-informed materials and teaching methods in the lecture.

**C. The students**

**1. Calculus-based Courses**

**a. Large lecture sections of the calculus-based course**

The calculus-based course consists primarily of engineering and computer science majors. The number of students registered for CBI, the first course in the sequence, each semester, is usually around 500, split among three lecture instructors. The number of students in the second course in the sequence, CBII, is around 400, split among two or three instructors. The instruction is primarily traditional lecture, with one one-hour recitation section and one two-hour lab, as described above. The labs and recitations are common among the three instructors each semester. Students from each of the lecture instructors are mixed in the labs and recitations.

**b. Honors section**

There is one honors section of the calculus-based class that is taken by students in the TTU Honors College and by some of the physics majors. It is usually a small class, consisting of 10 – 24 students. Sometimes the honors students take the same laboratories as the large lecture sections and sometimes they do not, depending on the instructor. The results from some honors sections are included for completeness.

**2. Algebra-based Courses**

**a. Large lecture sections of the algebra-based course**

The algebra-based class consists mostly of pre-health science majors, including pre-medical, pre-dental, pre-physical therapy, etc. The number of students registered each semester in the first course in the sequence is usually around 250-300 and has been around 100-150 in the second course in the sequence in recent semesters.[27] Except for the inquiry-based section of the course, the students are divided into two lecture sections taught by two lecture instructors each semester. The instruction is primarily traditional lecture, with one one-hour recitation section and one two-hour lab each week. The labs and recitations are common among the three instructors. Students from each of the lecture instructors are mixed in the labs and recitations.

**b. Inquiry-based, laboratory-based section**

In addition to the large lecture classes, we teach an inquiry-based, laboratory-based section of the algebra-based course every semester that was developed with National Science Foundation (NSF) support[15-16] starting in 2001. The course was developed specifically for health science majors, taking their needs, learning styles, backgrounds and motivations into account. It is taught without a text in a Workshop-Physics style[19] environment and is an inquiry-based course in the manner of Physics by Inquiry,[20] developed by the Physics Education Group at the University of Washington, but at the algebra-based level. The materials were developed by modifying and adapting parts of existing materials designed for other populations and integrating them with new units in our own format, creating a course aimed specifically at health science majors.

The curriculum was designed to be taught in a laboratory-based environment with no lecture and no text; however, a text can be used. Students work through the units in groups, learning about the world around them through experimentation, learning to develop both quantitative and qualitative models based on their observations and inferences. The materials consist of the laboratory units, pretests, readings and exercises. There are also homework sets, exams and quizzes. The students sign up on a first-come, first-serve basis.

The free-response pre- and post-tests, while designed for the lecture-based classes, were administered to the students in this section of the algebra-based course, also.

## III. DEVELOPMENT AND ADMINISTRATION OF ASSESSMENT INSTRUMENTS

The assessment instruments were a set of free response pre- and post-tests written for the large lecture sections of both the algebra-based and calculus-based courses by three of the researchers on the NIH grant. Two sets of free response pre- and post-tests were developed, one for the first course in the sequence and one for the second course in the sequence for both the algebra and calculus-based classes. The problems in the first course consisted mostly of mechanics problems and the problems in the second half of the course consisted mostly of electricity and magnetism problems, although there were a few pretests and quizzes on other topics in each of the semesters. When possible, we tried to administer the same problems in both the calculus- and algebra-based classes. This allowed for comparison between the two classes, which is actually very interesting in our environment. The pre-tests were designed to assess the knowledge of the students on either the lab or recitation content for that week. The quizzes were designed to match one of the pre-tests, if possible. They were designed to assess content or skills covered in the lab, recitation, and/or lecture. Both the pre-tests and the quizzes were free response and required the students to show their work and explain their reasoning. We have chosen, as examples, to present four matching pre- and post-tests from the second semester courses, ABII and CBII. The matching pre- and post-tests selected and examples of acceptable answers are given in Appendix II.

In both ABII and CBII, a pre-test was administered at the beginning of every lab and a quiz, which served as the post-test, was administered bi-weekly in the recitation sections. The pre-tests were collected and scanned for research. Pre-tests were not graded, but were counted towards a participation grade and the participation grade credit was not given, if the answer did not appear to be a serious attempt at solving the problem or answering a question. Each week that a quiz was given, the quizzes were scanned and saved for research before being graded by the TAs and returned to the students. The grading rubric used by the TAs was not necessarily the analysis rubric used by the researchers, which is given in Appendix III.

## IV. ANALYSIS

### A. Pre-test, post-test scores and common types of answers

The scanned pre- and post-tests were analyzed by researchers using an analysis rubric to categorize the answers based on correct choices and explanations. The most common answer categories used were: Completely correct (CC), Correct choice, partially correct explanation (CP), Correct choice, incorrect explanation (CI), and Incorrect (I). However, in some problems, the categories were simply Correct (C), Partially correct (P) and Incorrect (I). The researchers also noted common incorrect answers and solution processes, as well as common types of answers, and recorded these. The answer categorization rubric for each of the pre- and post-tests is given in Appendix III and samples of common incorrect answers and explanations are given in Appendix IV.

### B. Gain

In addition to reporting the post-test results as the percentage of students in each of the four categories CC, CP, CI and I, we report the gain, the change of an individual student from pre-test to post-test. The gain is determined by comparing the movement between categories for each student. Given the four categories CC, CP, CI, I, if a student moves two steps upward, from I or CI to CC, the student is awarded a gain value of 2 points. If a student moves one step upwards, from CI or I to CP or CP to CC, the student is given a gain value of 1 point. If the student moves one step down, CC to CP or CP to CI or I, the student is given a gain value of -1. Similarly, movement from CC to I or CI, is awarded -2 points. The student then has a gain score in the range of -2 to 2. In addition, there are two ways the student can have no gain. If the student has I or CI on the pre-test and still has I or CI on the post-test, the student gain is reported as 0. However, if the student has CC on the pre-test and still has CC on the post-test, the gain is reported as 0+. This student has no gain, but understands the concept or skill being assessed. A student then has a gain score in the range of -2 to 2, with the possibility of 0 or 0+.

The gain is then plotted as six possible categories: 1) complete decrease (CD), gain value -2 points, 2) partial decrease (PD), gain value -1 point, 3) no gain (NG), gain value 0 points, 4) no gain positive (NGP), gain value 0+ points, 5) partial gain (PG), gain value 1 point, 6) complete gain (CG), gain value 2 points. We present both gain and post-test plots in the Results in Figures 1 – 8.

## V. RESULTS

We present the results of the written pre- and post-testing for four examples of matched pre- and post-tests used in our study. All are on concepts or skills in the electricity and magnetism part of the course and they represent results at different points during the semester. We present results on the same pre- and post-tests in the algebra- and calculus-based classes and across lecture and lab teaching styles and pedagogy.

### A. Problem 1 Electrostatics

The first problem is a problem on two charged parallel plates, given early in the semester when students are studying electrostatics. The problem is given in Appendix II. In between the pre- and the post- test, the students in the large lecture sections, ABII and CBII, have done an experiment in lab that consists of two 2-D charged conducting plates on conducting paper of essentially the same set-up as in the problem. They experimentally determined the location of equipotential lines and the direction of the electric field based on measurements with a voltmeter and used those measurements to calculate the magnitude of the electric field. The inquiry-based students did not work through that particular lab, but had similar questions as part of thought experiments in their lab. The problem has three parts. The first part is a conceptual question, the second part is a laboratory question and the third part requires understanding the meaning of an equation.

### i. Post-test

The results are given in Figures 1 and 2. Figure 1 has the post-test results for each part by class, lab style, and teaching style. Figure 2 has the gain for each part of the problem by class, lab style, and teaching style. The number of students, N, whose matched pre- and post-tests were analyzed is also given.

The Calculus-based classes with traditional labs (T) and traditional lecture (TL) teaching style had the lowest, practically zero, post-test and gain scores. No scores are particularly high, but it is only with both PER labs and PERL instruction in the large algebra-based sections and in the Inquiry class that we get CC above 40% on part (a), and only in the Inquiry course is CC above 20% in parts (b) and (c). Also the inquiry-based students outperform the honors students in all parts of the problem. The considerable percentage of CP answers in part (b) is due to a correct explanation that does not explicitly identify the potential difference measurement to be made in the direction of the electric field or between equipotential lines, as explained in Appendix II.

It is also interesting to note the percentage of students that chose the correct answer in part (c), but were not able to correctly explain their reasoning. We saw a relatively high percentage of CI answers (often higher than in this problem) across many problems in our study, which prompted a further study on the effect of problem format on students' answers. It is reported in a separate paper.[14]

**ii Gain**

The gain is a nice way of plotting the transition from pre-test to post-test, as it records the individual student's movement from one category to another. Complete gain (CG) is achieved by an incorrect pre-test answer, CI or I, to a completely correct, CC, post-test answer. Partial gain (PG) is a move from completely incorrect, CI or I, to partially correct, CP, or a move from partially correct, CP, to completely correct, CC. With a gain calculation, it is also possible to see that students can sometimes answer the pre-test correct or partially correct, but the post-test incorrectly. This is counted as partial or complete decrease (PD or CD) and does happen sometimes in practically all of the classes. The no gain categories are straight forward, with no gain (NG) meaning that the concepts or skills were not learned in between the pre-test and the post-test based on this assessment. However, the no gain positive (NGP) category is important, because it indicates students who already understood the concept or skill before the pre-test and still understand it when they take the post-test. So there are three positive categories (CG, PG and NGP) and three negative categories (CD, PD and NG) represented in the gain plot. It is useful to look at CG, PG and NGP to see the percentage of students that understand, at least partially, the skills or content knowledge assessed.

Looking at CG, PG and NGP together, the algebra-based PER/PERL large lecture class, the inquiry class and the honors section are at least at 50% completely or partially correct understanding on part (a) and the inquiry-based section and the honors section are above 60% completely or partially correct understanding on part (b).

**iii. Answer types**

Common incorrect responses are given in Appendix IV. The most common error in part (a) was to rank the electric field magnitude by the proximity to one of the plates. Students either ranked the field magnitude as higher closer to the positive plate or higher closer to the negative plate. Over 70% of the incorrect answers fell into this category.

Part (b) incorrect answers were characterized by vagueness, lack of detail and incompleteness. The students did not have the ability to identify and verbalize the key measurements they had made in the previous week's lab and write a clear and concise description of the measurements and how to calculate the magnitude of the electric field. These responses indicate that our students are not able to explain the measurements they took in a previous week's lab and not able to explain how to use those measurements to determine the value of a specific quantity (in this case the electric field). This question gave us very important information on what our students are getting (or not getting) out of the laboratory experience and their ability to connect it to the physics they are learning in lecture and recitation sections.

The most common incorrect answer to part (c) was to choose (ii) and argue that E = ΔV/d was the equation learned in class. This gives us information on the ability of students to understand equations and how to apply them.

### B. Problem 2 Electric Circuits

Pre- and post-tests on electric circuits varied considerably throughout our study. We report two parts of a problem that we have post-test data for in many different courses of different lab and teaching styles. We report the gain on one of those problems, as that is the only electric circuit problem gain that we have across all class, lab and lecture styles. We report the gain only for classes that we have matching pre- and post- test data on that problem. The post-test data is presented in Figure 3 and the gain data is in Figure 4. The gain data matches Pre-test Problem 2b to Post-test Problem 2a.

On this problem we also have data from one semester in the large lecture section algebra-based course that used mostly material from Real Time Physics[21] in the laboratories. While Real Time Physics is also PER-informed material, we label it RT and separate it out from locally written PER-informed materials (PER).

#### i. Post-test

Students, in general, did better on this problem than on Problem 1. Four groups of students, the Inquiry-based class, the calculus-based class with PER labs and PERL instruction, the algebra-based class with RT labs and TL instruction and the honors class all had CC scores between 40% and 60% on part (a). The Inquiry-based class and the algebra-based class with RT labs and TL instruction also had scores roughly in that range on part (b). Again, the algebra-based, inquiry-based class outperforms all of the other classes, including the honors students. And also again, it is interesting to note the percentage of students that chose the correct answer, but were not able to correctly explain their reasoning.

This problem was given later in the semester than Problem 1 and is the one problem where the calculus-based students with the same teaching style outperform the algebra-based students. It is interesting that it is on an electric circuits problem, as the calculus-based students are predominantly engineering students and may have been familiar with circuits before they took the course. However, it is very interesting that the algebra-based students at our institution often outperform the calculus-based students both on free-response testing, as reported in this paper, and on conceptual inventories, as reported elsewhere.[13] We are undertaking further study on this issue.

**ii. Gain**

The gain is given for those classes for which we have a matching pre- and post-tests. It reflects the same trends, with the INQ class, the calculus-based PER/PERL class and the honors class, in this case, with the most gain. The calculus-based courses have a higher percentage of NGP, indicating more correct knowledge of the content before the pre-test was taken. The calculus-based students may have learned this particular content before taking the course, which is highly probable for the engineering students in the calculus-based courses and probably less likely for the pre-health science students in the algebra-based courses.

**iii. Answer types**

Common incorrect responses are given in Appendix IV. The most common incorrect answer was to argue that the total resistance increases when a resistor is added and that the current through the battery therefore decreases. The next most common argument was that the current remains constant, independent of the number of resistors. Both of these conceptions have been well documented and reported in the literature.[28-29]

The most common incorrect answer in part (b) was that the voltage remains the same because it is independent of the resistors connected. This has also been well documented.[28-29]

**C. Problem 3 Magnetism**

The third problem we present assesses students' ability to interpret graphs and analyze data. The pre-test is not a direct match to the post-test, so we present the results of the pre- and post-tests, instead of the post-test and gain. However, the number of students in the CC category on the pre-test was essentially zero, so the post-test score is essentially the gain score. The pre-test and post-test results are presented in Figures 5 and 6.

**i. Pre-test**

The pre-test assesses students' ability to interpret a graph of the magnitude of the magnetic field as a function of the distance from a current-carrying wire. It assesses their ability to match a graph to an equation. The Inquiry-based classes did not take this pre-test. None of the students in the large lecture sections did very well on the pre-test problem.

Common incorrect responses to the pre-test are given in Appendix IV. The most common incorrect explanations were to explain choices of graphs B or C as being correct because they described the exponential decay of the magnetic field. Students who chose the correct answer, B, were not able to correctly explain it and usually used an exponential argument. The most common CI error was to choose graph B and describe it as an exponential decrease that is not as fast a decrease as graph C.

The second most common error was to argue a linear decrease because the field got weaker the further away you were from the wire. Many students wrote down the correct relationship or correct formula for the magnetic field and the distance from the wire and then described the function as linear.

Again we see a relatively high percentage of false positives (CI answers), students who could choose the correct answer but not correctly explain it.

**ii. Post-test**

The post-test assesses students' ability to distinguish two sets of data, one for an electric field and one for a magnetic field. Between the pre- and the post-test, the students in the large lecture classes had done a laboratory in which they measured and graphed the magnitude of the magnetic field at different distances from a current carrying wire. The Inquiry class had not done a quantitative measurement of that type at the time the assessment was given. However, they had experimentally qualitatively analyzed the magnitude of a magnetic field at different distances from a current-carrying wire and studied the equations involved. The intent of the problem was that students would use graphing skills and describe how to distinguish the two sets of data by graphing them and analyzing the curves, although there are other correct ways to work the problem.

None of the classes did particularly well on the post-test, with the highest post-test (which is essentially the gain) being about 25%. Both the pre- and the post-test gave us significant information on our students' ability to interpret graphs and analyze data. Our students are not very proficient at those skills, based on this particular assessment.

This is the one problem on which the inquiry-based students did not do as well. We will be assessing the teaching of graphing skills in all of our classes and also doing further testing on the students abilities in this area.

**iii. Answer types**

Common errors are given in Appendix IV. One of the most common errors was a lack of detail. In particular, students would argue that you simply plug the numbers into the formula and see which data fit. This does not address the fact that you don't know the current, I, or the charge, Q, and you need to take a ratio or determine the constant in order to determine how the field data decreases with distance.

Another common incorrect answer was to interpret the question as asking the student to determine which set of data is from measurements of an electric field and which is from measurements of a magnetic field, not the process of how you would determine it. Students made arguments, even though no units are given, that either the electric field or the magnetic field would produce larger values. These students misunderstood the problem as trying to determine which was the magnetic field and which was the electric field and did not focus on the process of how you would determine it.

Another common answer was that you should re-do the experiment, which also did not indicate an understanding of the purpose of the problem.

**D. Problem 4 Induction**

Problem 4 assesses the students' understanding and ability to use Faraday's Law and Lenz's Law. Between the pre- and the post-test, the students in the large lecture sections had worked through a laboratory (Appendix I) in which they had performed both qualitative experiments demonstrating Faraday's and Lenz's Law and experiments in which they used measurements of a changing magnetic field and the current through a solenoid to calculate the average emf through the solenoid two different ways. The inquiry-based class had done qualitative experiments that demonstrated Faraday's and Lenz's Law and studied the mathematical use of the equations to solve problems. The post-test and gain graphs are given in Figures 7 and 8. The gain is given for those classes for which we have both the pre- and post-test data.

**i. Post-test**

Here again, the inquiry-based students outperform the students in the other sections, the algebra-based students outperform the calculus-based students and the traditionally taught students with traditional labs perform the least well. This problem was administered near the end of the semester. The inquiry-based students had a high percentage of CC answers on this problem, with about 75% completely correct on part (a) and about 50% completely correct on part (b). The algebra-based classes and the PER/PERL calculus-based class had between 30% and 50% CC answers on part (a). The algebra-based classes were also above 20% CC on part b.

**ii. Gain**

The gain scores reflect the same trend as the post-test scores. The algebra-based classes have a higher percentage of NGP, indicating that they already knew the concept before the pre-test.

**iii. Answer types**

Common incorrect answers are given in Appendix IV. Many of the incorrect answers in part (a) reflected a conception that the loop simply had to be in, or partially in, the field but not that the flux had to be changing for there to be current. Some students thought the location of the ammeter determined whether current flowed, i.e., if the ammeter was inside the field, there would be current.

In part (b), many students used the right hand rule incorrectly. Some students did not understand that the external magnetic field was not produced by the currents in the loops.

In part (c), the answers varied widely; there were not common incorrect answers.

## VI. DISCUSSION AND CONCLUSIONS

In general, the students in the inquiry-based course performed at least as well as, but usually better, than all of the other students, in both the calculus-based and algebra-based classes, including the calculus-based honors sections. The traditionally taught calculus-based classes (T/TL) demonstrated the lowest performance on all problems. The introduction of PER instruction in the laboratories only or in both the lecture and lab did not have much effect at the beginning of the semester, but did increase the performance somewhat later in the semester, with the PER/PERL instruction resulting in a larger gain than the PER/TL instruction. The algebra-based classes consistently scored higher than the calculus-based classes, except on Problem 2, with the PER/PERL instruction sometimes coming close to the results of the inquiry-based course. We have an untested hypothesis on the performance of the algebra-based students compared to the calculus-based students. Our calculus-based engineering students only need a C in physics to continue in engineering, but our algebra-based health science students usually need an A in physics for entrance into professional schools in fields such as medicine, dentistry and physical therapy, to name a few. We think this effects students' motivation and is a factor in the performance of our students on these assessments, but this needs further study.

We also notice that many of the results of these post-tests are not particularly high. The percentage of correct answers on any of the questions is rarely above 30% in any of the calculus-based classes. The PER/PERL algebra-based students, the inquiry-based students and the honors students, on the other hand, see the percentage of correct answers at least in the 40 – 60% range on some parts of some problems.

Part of the reason for this study was to introduce assessment instruments beyond conceptual inventories into our assessment and evaluation of our courses. It is interesting to compare these results to the results of conceptual inventories reported in a previous paper.[13] In that study, as shown in Table 1, the average normalized gain on the Brief Electricity and Magnetism Assessment (BEMA)[8] of the PER/PERL calculus-based, PER/PERL algebra-based and INQ students was 0.19 +/- 0.01, 0.17 +/- 0.02 and 0.16 +/- 0.02, respectively.  The latter two were not significantly different. The average BEMA post-test scores were 39.0 +/ 1.0, 35.7 +/- 1.4 and 34.7 +/- 1.7 for the PER/PERL calculus-based, PER/PERL algebra-based and INQ students, respectively. The PER/PERL calculus-based post-test and gain scores are significantly different from the post-test and gain scores of the others (at p < 0.05 by a Student's T-test), but they are not much higher. However, this study, even though it is a series of "snapshots" of students' understanding, gives evidence that the inquiry-based students perform better at free-response questions that require students to show their work and explain their reasoning. One can argue this requires different skills than answering conceptual questions in a multiple-choice format. It serves as evidence that there is a need for more comprehensive assessment instruments that go beyond conceptual inventories that can be compared across classes and universities that assess different aspects of instruction, including problem solving, laboratory skills, aspects of critical thinking, and modeling, in addition to conceptual understanding, in order to give us a more complete picture of our instructional methods.

**VII. ACKNOWLWDGEMENTS**

We thank the National Institutes of Health (NIH) for their support of this project, for the funding of NIH Challenge grant #1RC1GM090897-02. Any opinions, findings and conclusions or recommendations expressed are those of the authors and do not necessarily reflect the views of the NIH.

| Lab/Lecture Style | N | %Lab/Rec. | Pretest | S.E. | Posttest | S.E. | g | S.E. |
|---|---|---|---|---|---|---|---|---|
| Alg.-based | | | | | | | | |
| RT/TL | 192 | 20% | 23.6 | 0.5 | 28.9 | 0.7 | 0.07 | 0.01 |
| PER/TL | 64 | 20% | 22.7 | 0.9 | 31.7 | 1.4 | 0.11 | 0.02 |
| PER/PERL | 58 | 20% | 21.9 | 0.9 | 34.7 | 1.7 | 0.17 | 0.02 |
| INQ | 62 | N/A | 20.7 | 1.0 | 35.7 | 1.4 | 0.16 | 0.02 |
| Calc.-based | | | | | | | | |
| T/TL | 345 | 10% | 21.9 | 0.4 | 27.5 | 0.6 | 0.07 | 0.01 |
| PER/TL | 241 | 10% | 21.7 | 0.5 | 30.4 | 0.7 | 0.08 | 0.01 |
| PER.TL | 105 | 30% | 23.84 | 0.9 | 33.7 | 1.3 | 0.13 | 0.01 |
| PER/PERL | 200 | 25% | 24.3 | 0.6 | 39.0 | 1.0 | 0.19 | 0.01 |
| H PER/TL | 9 | 20% | 25.2 | 2.5 | 43.3 | 2.3 | 0.25 | 0.05 |

Table 1. Results for BEMA for algebra-based and calculus-based courses by lab and teaching style. Lab styles are labeled by traditional (T), Real Time Physics (RT), and PER-informed (PER). The teaching styles are labeled by traditional (TL), PER-informed (PERL) and Inquiry-based (INQ). Honors sections are labeled with an *H*.

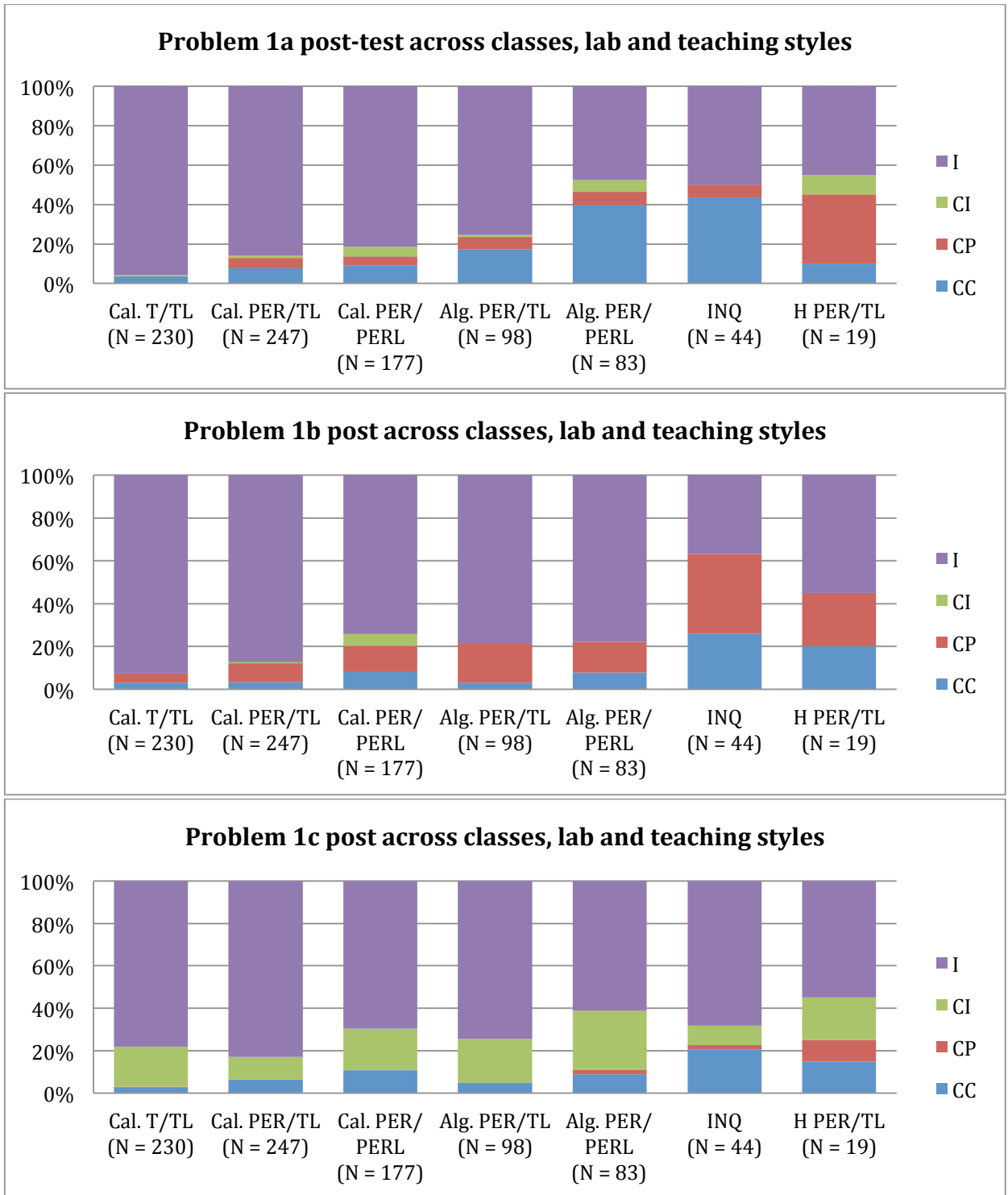

Figure 1: Problem 1 post-test scores across classes, lab and teaching styles. Answers are categorized as Completely correct (CC), Correct answer, partially correct explanation (CP), Correct choice, incorrect explanation (CI), or Incorrect (I). Laboratory style is categorized by traditional (T) and PER-informed (PER) and teaching style is categorized by traditional (TL), PER-informed (PERL) and Inquiry-based (INQ). Honors sections are indicated with an H.

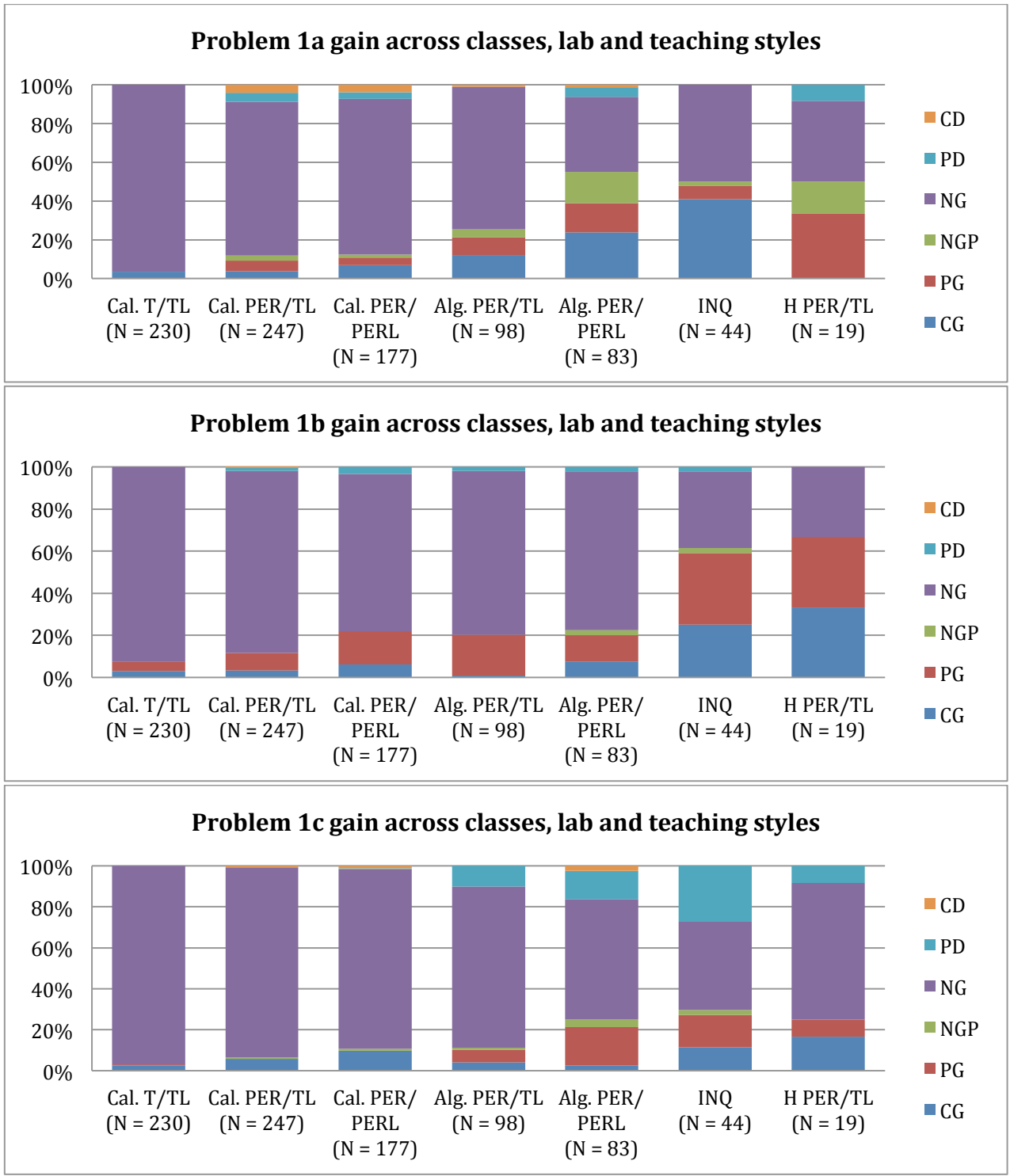

Figure 2: Problem 1 gain values across classes, lab and teaching styles. Gain is characterized as: complete decrease (CD), partial decrease (PD), no gain (NG), no gain positive (NGP), partial gain (PG) or complete gain (CG). Laboratory style is categorized by traditional (T) and PER-informed (PER) and teaching style is categorized by traditional (TL), PER-informed (PERL) and Inquiry-based (INQ). Honors sections are indicated with an H.

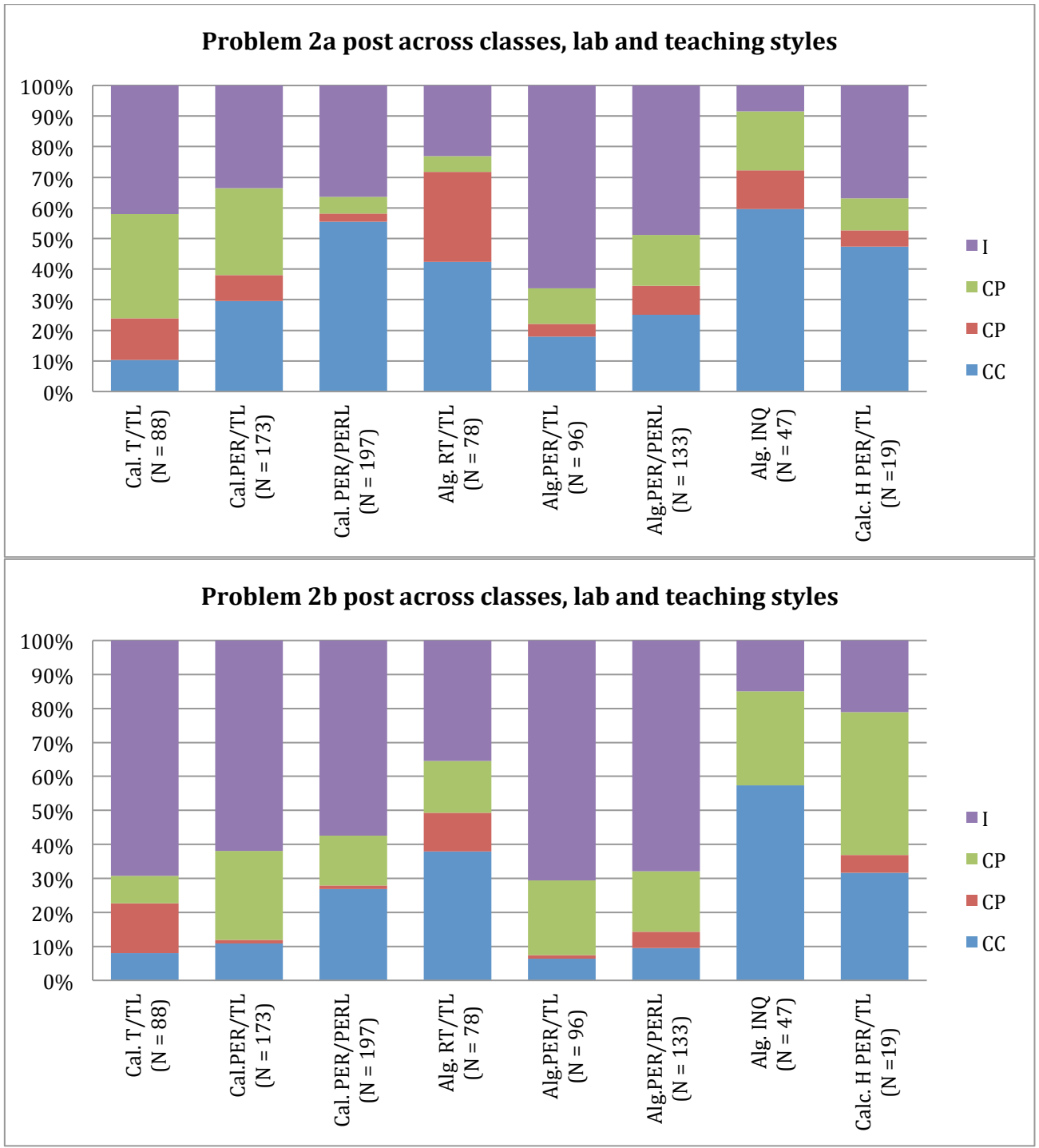

Figure 3: Problem 2 post-test scores across classes, lab and teaching styles. Answers are categorized as Completely correct (CC), Correct answer, partially correct explanation (CP), Correct choice, incorrect explanation (CI), or Incorrect (I). Laboratory style is categorized by traditional (T), PER-informed (PER) and Real Time Physics (RT) and teaching style is categorized by traditional (TL), PER-informed (PERL) and Inquiry-based (INQ). Honors sections are indicated with an H.

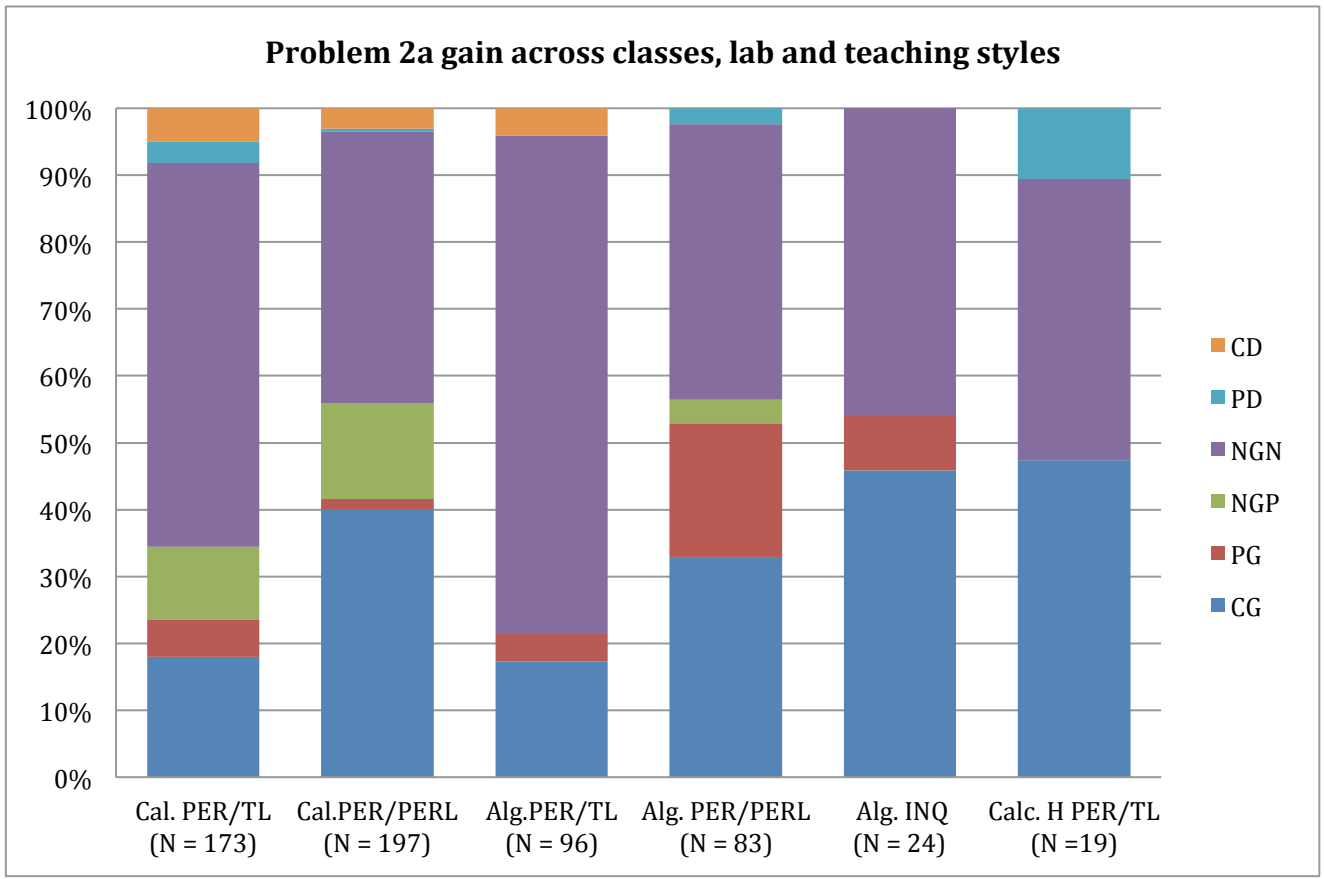

Figure 4: Problem 2a gain values across classes, lab and teaching styles. Gain is characterized as: complete decrease (CD), partial decrease (PD), no gain (NG), no gain positive (NGP), partial gain (PG) or complete gain (CG). Laboratory style is categorized by traditional (T), PER-informed (PER) and Real Time Physics (RT) and teaching style is categorized by traditional (TL), PER-informed (PERL) and Inquiry-based (INQ). Honors sections are indicated with an H.

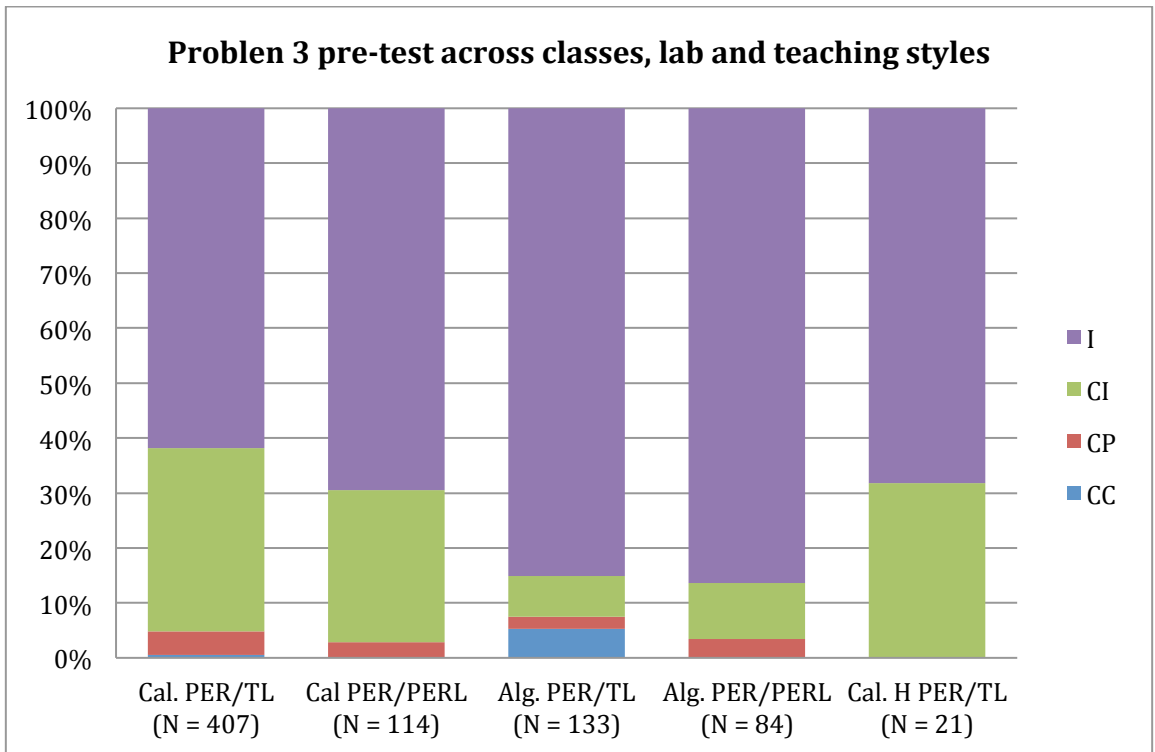

Figure 5: Problem 3 pre-test scores across classes, lab and teaching styles. Answers are categorized as Completely correct (CC), Correct answer, partially correct explanation (CP), Correct choice, incorrect explanation (CI), or Incorrect (I). Laboratory style is categorized by traditional (T), PER-informed (PER) and Real Time Physics (RT) and teaching style is categorized by traditional (TL), PER-informed (PERL) and Inquiry-based (INQ). Honors sections are indicated with an H.

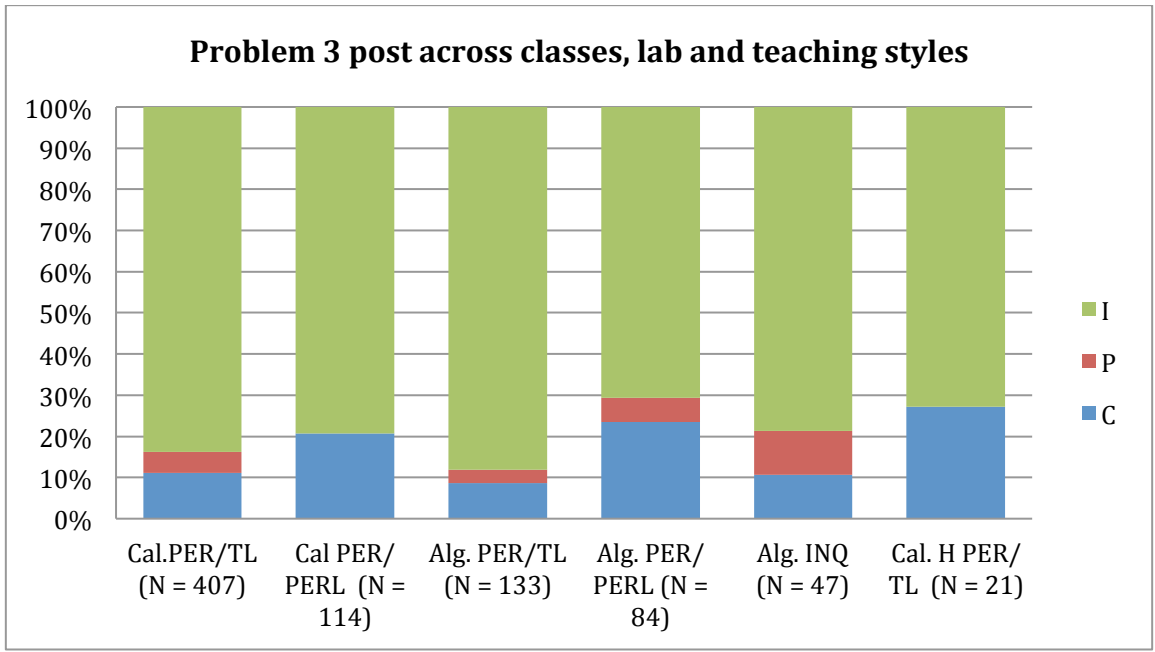

Figure 6: Problem 3 post-test across classes, lab and teaching styles. Answers are categorized as Correct (C), Partially correct (P), or Incorrect (I). Laboratory style is categorized by traditional (T), PER-informed (PER) and Real Time Physics (RT) and teaching style is categorized by traditional (TL), PER-informed (PERL) and Inquiry-based (INQ). Honors sections are indicated with an H.

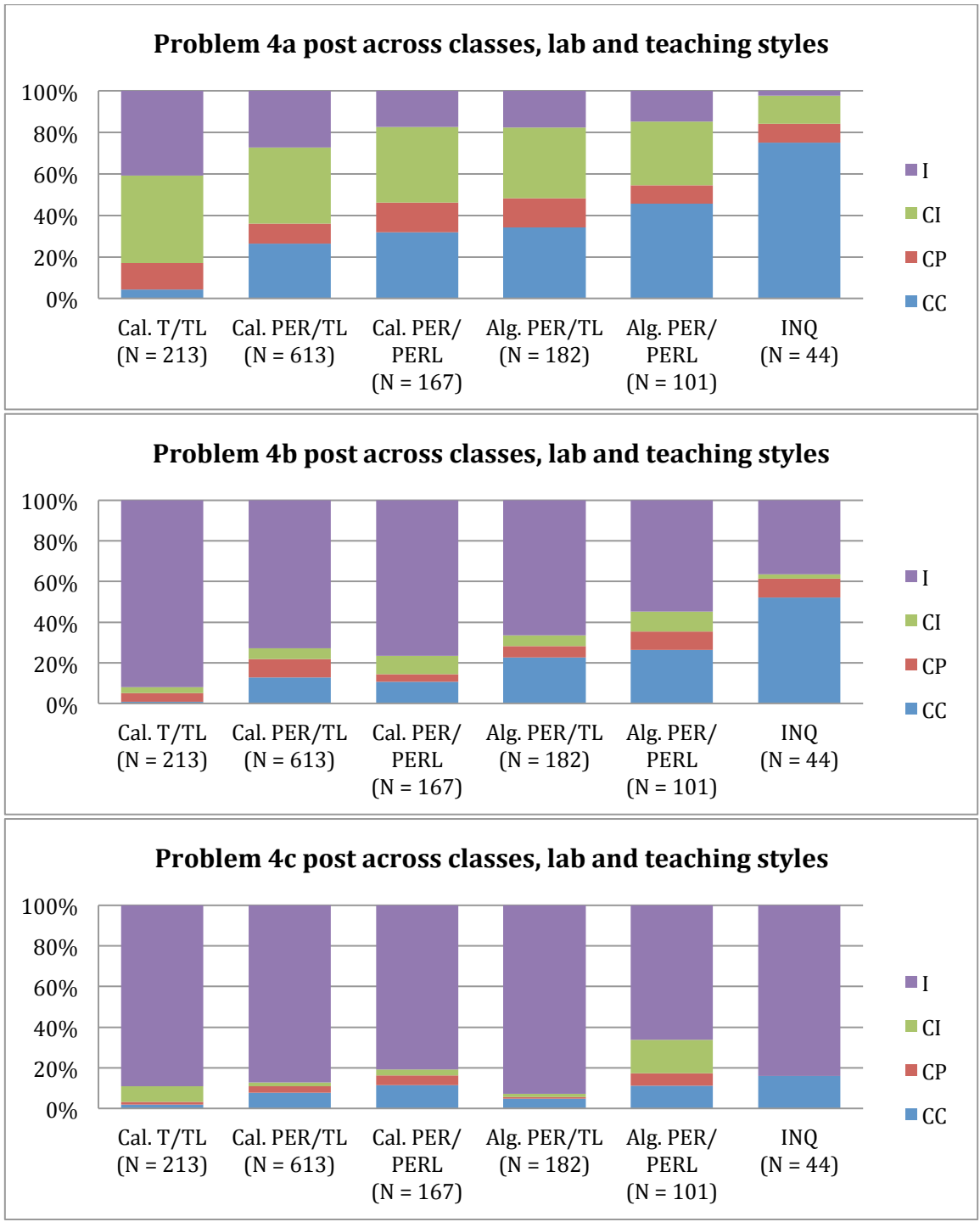

Figure 7: Problem 4 post-test scores across classes, lab and teaching styles. Answers are categorized as Completely correct (C/C), Correct answer, partially correct explanation (C/P), Correct choice, incorrect explanation (C/I), or Incorrect (I). Laboratory style is categorized by traditional (T), PER-informed (PER) and Real Time Physics (RT) and teaching style is categorized by traditional (TL), PER-informed (PERL) and Inquiry-based (INQ). Honors sections are indicated with an H.

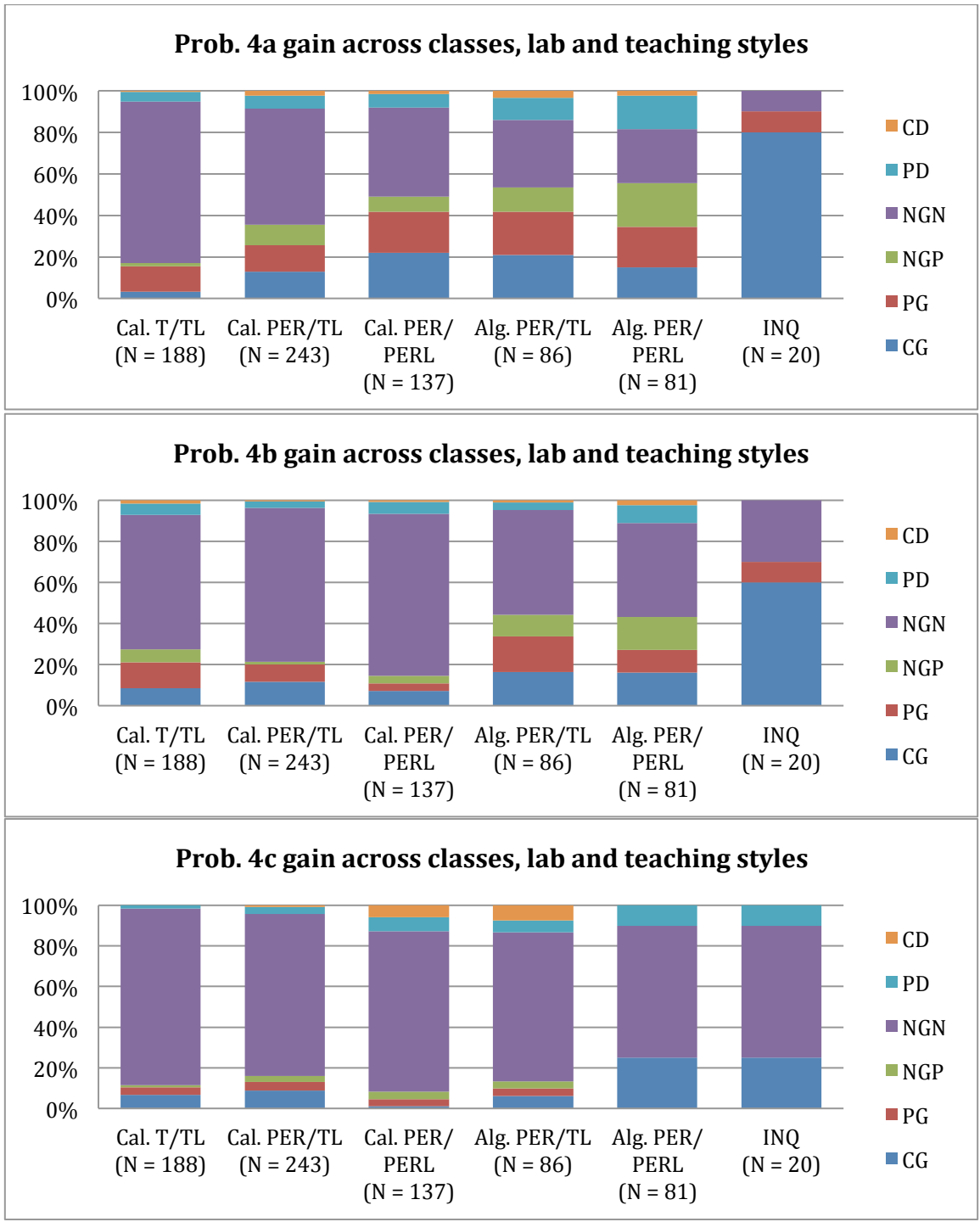

Figure 8: Problem 4 gain values across classes, lab and teaching styles. Gain is characterized as: complete decrease (CD), partial decrease (PD), no gain (NG), no gain positive (NGP), partial gain (PG) or complete gain (CG). Laboratory style is categorized by traditional (T), PER-informed (PER) and Real Time Physics (RT) and teaching style is categorized by traditional (TL), PER-informed (PERL) and Inquiry-based (INQ). Honors sections are indicated with an H.

# References

1. See for example, the Assessment Instrument Information Page, Physics education Research Group, North Carolina State University, http://www.ncsu.edu/per/TestInfo.html, 10/09/13.

2. Julia I. Smith and Kimberly Tanner, "The Problem of Revealing How Students Think: Concept Inventories and Beyond," *CBE Life Sci. Educ.* vol. 9, no. 1, pp. 1-5.

3. Marcos D. Caballero, Edwin F. Greco, Eric R. Murray, Keith R. Bujak, M. Jackson Marr, Richard Catrambone, Matthew A. Kohlmyer, and Michael F. Schatz, "Comparing large lecture mechanics curricula using the Force Concept Inventory: A five thousand student study," *Am. J. Phys.* **80**, (7), pp. 638 - 644.

4. The results of Physics Education Research (PER) are not always introduced into curricula by taking one set of materials developed and implementing only that set of materials in the exact manner as the developers intended. It is common for instructors to adopt pieces of curricula or materials or instructional strategies and interweave those pieces into their own course. Sets of materials have also been developed locally by PER practitioners based on their knowledge of the field (their knowledge of the research based on talks and workshops presented at national meetings, their own research and the literature). We operationally define PER-informed instructional materials and pedagogy as those developed by instructors well-informed and well-read in PER or researchers in PER that may include pieces of published PER-based curricula, materials or instructional strategies, problems from those materials or from PER-based texts or literature and PER-based strategies that have been demonstrated to be effective in some context. We use the term PER-informed because it is not the adoption of a single PER-based curricula or instructional method, but a compilation of materials locally by instructors well-read and informed in the use of various PER-based curricula, materials and instructional strategies.

5. National Institutes of Health (NIH) Challenge grant #5RC1GM090897-02, "An Assessment of Multimodal Physics Lab Intervention Efficacy in STEM Education," to assess for interventions to the laboratory curriculum at Texas Tech University, PI's Beth Thacker and Kelvin Cheng.

6. We are not aware of any comprehensive assessment instruments for the undergraduate introductory physics courses, although instruments do exist for other populations, such as Advanced Placement (AP) physics exams for high school students. We are interested in the development of such an examination, but designed for the purpose of providing feedback to instructors about their students' performance and for use in assessing changes to courses and curricula.

7. I. Halloun and D. Hestenes, "The Initial Knowledge State of College Physics Students," *Am. J. Phys.* **53**, 1043-1055 (1985).

8. L. Ding, R. Chabay, B. Sherwood, and R. Beichner, "Evaluating an electricity and magnetism assessment tool: Brief electricity and magnetism assessment," Phys. Rev. ST Phys. Educ. Res. 2, 010105 (2006).

9. D. Hestenes and M. Wells, A Mechanics Baseline Test, *The Physics Teacher* **30**: 159-165 (1992).

10. D. Maloney, T. O'Kuma, C. Hieggelke, and A. Van Heuvelen, "Surveying students' conceptual knowledge of electricity and magnetism", Am. J. Phys. 69, S12 (2001).

11. In particular, we used the Colorado Learning Attitudes about Science Survey (CLASS), W. K. Adams, K. K. Perkins, N. S. Podolefsky, M. Dubson, N. D. Finkelstein, and C. E. Wieman, "New instrument for measuring student beliefs about physics and learning physics: The Colorado Learning Attitudes about Science Survey", Phys. Rev. ST Phys. Educ. Res. 2, 010101 (2006) and The Classroom Test of Scientific Reasoning (CTSR), Anton E. Lawson, "The development and validation of a classroom test of formal reasoning," J. Res. Sci. Teach. 15, 11 (1978).

12. We used the Reformed Teaching Observation Protocol (RTOP), MacIsaac, D.L. & Falconer, K.A. (2002, November). Reforming physics education via RTOP. The Physics Teacher 40(8), 479-485. for this purpose.

13. Beth Thacker, Hani Dulli, Dave Pattillo, and Keith West, "Lessons from a Large-Scale Assessment: Results from Conceptual Inventories," submitted to the American Journal of Physics.

14. Beth Thacker, Ganesh Chapagain, Mark Ellermann, Dave Pattillo and Keith West, "The Effect of Problem Format on Students' Responses," submitted to the American Journal of Physics.

15. National Science Foundation - Course, Curriculum and Laboratory Improvement grant CCLI #9981031, "Workshop Physics with Health Science Applications" for development of materials for the introductory, algebra-based physics course with a health science based content focus in an interactive, computer-based laboratory setting, PI's Beth Thacker and Anne Marie Eligon.

16. National Science Foundation - Course, Curriculum and Laboratory Improvement grant CCLI-EMD #0088780, "Humanized Physics -- Reforming Physics Using Multimedia and Mathematical Modeling" for development of materials for the introductory, algebra-based physics course with health science based content using mathematical modeling and multimedia, co-PI's Robert G. Fuller and Vicki

L. Plano Clark, Beth Thacker, Nancy L. Beverly, Chris D. Wentworth, and Mark W. Plano Clark.

17. Beth Thacker, Abel Diaz, and Ann Marie Eligon, "The Development of an Inquiry-based Curriculum Specifically for the Introductory Algebra-based Physics Course," arXiv:physics/0702247.

18. Jennifer Wilhelm, Beth Thacker and Ronald Wilhelm, "Creating Constructivist Physics for Introductory University Classes," Electronic Journal of Science Education Vol. 11, No. 2, 19-37, (2007).

19. Priscilla W. Laws, *Workshop Physics Activity Guide*, (John Wiley & Sons, Inc., New York, 1999).

20. Lillian C. McDermott and the Physics Education Group, Physics by Inquiry Volumes I and II, (John Wiley and Sons, NY, 1996).

21. David R. Sokoloff, Ronald K. Thornton, Priscilla W. Laws, "Real Time Physics Active Learning Laboratories Modules 3 and 4, 2nd Edition, Wiley, (2004).

22. S. Lance Cooper, Bernard M. Dick, and Alexander Weissman, Physics 211 Laboratory experiments, Department of Physics, College of Engineering, University of Illinois at Urbana Champaign, Stipes Publishing, (2004).

23. C. J. Hieggelke, D. P. Maloney, T. L. O'Kuma, Steve Kanim, E&M TIPERs: Electricity & Magnetism Tasks, (Addison-Wesley, 2005).

24. T L O'Kuma, D P Maloney, C J Hieggelke, Ranking Task Exercises in Physics: Student Edition, (Addison-Wesley, 2003).

25. Arnold B. Arons, Teaching Introductory Physics, (Wiley, 1996).

26. Arnold B. Arons, A Guide to Introductory Physics Teaching, (Wiley, 1990).

27. We are presently studying the drop in the number of students from the first to the second semester of the algebra-based course. This drop is striking because most of the students taking the first semester need to take the second semester in order to meet the requirements of their major. We have noted an increased number of students taking the second semester course in the summer sessions and it is possible that some are taking the course at different institutions in the summer. This is still being researched.

28. L. C. McDermott and P. S. Shaffer, "Research as a guide for curriculum development: An example from introductory electricity. Part I: Investigation of student understanding," *Am. J. Phys.* **60** , 994 (1992); Erratum: *Am. J. Phys.* **61** , 81 (1993)

29. P. S. Shaffer and L. C. McDermott, "Research as a guide for curriculum development: An example from introductory electricity. Part II: Design of instructional strategies," *Am. J. Phys.* **60** , 1003 (1992).

# Appendix I

# LABORATORY 9
# MAGNETISM III: FARADAY'S LAW, LENZ'S LAW, INDUCTION

## Objectives

- to observe that a changing magnetic field can give rise to a current

- to be able to explain the concept of magnetic flux

- to be able to explain the concept of electromotive force ($\xi$) or emf

- to be able to determine by measurement or calculation the electromotive force ($\xi$) or emf of a changing magnetic flux

- to be able to determine the magnitude and direction of a current set up by a changing magnetic field (Lenz's Law)

**Overview:** In this lab, we will explore Faraday's Law and Lenz's law.

**Equipment:**
    1 cow magnet
    1 solenoid
    1 small compass
    1 analog ammeter

**Exploration 1: Lenz's Law**

**Exploraton 1.1** Take the pre-test for this lab.

**Exploration 1.2** Connect the solenoid to an analog ammeter, as shown in the diagram below.

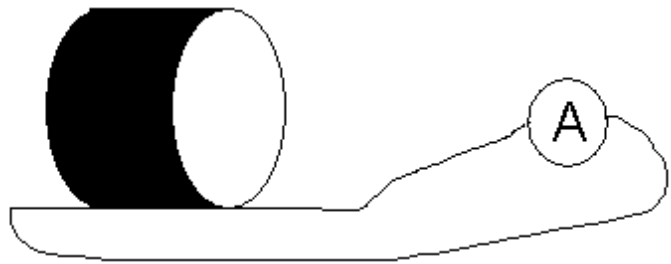

Figure 9: Solenoid and ammeter connection for Laboratory Exploration 1.2.

**a.** Determine which end of the cow magnet is north. Take a cow magnet and move it back and forth near the solenoid. Observe the ammeter while the magnet is moving.

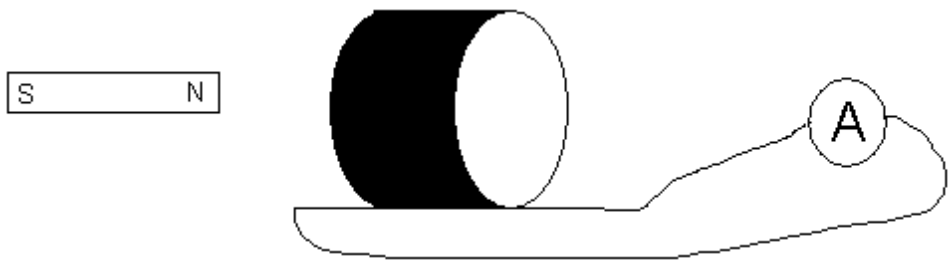

Figure 10: Set up with solenoid and cow magnet for Laboratory Exploration 1.2.

Put the north pole toward the loop. Then put the south pole toward the loop. Does the ammeter record a current? Which direction does the ammeter needle move when the north pole moves towards the magnet? Which direction does it move when the north pole moves away from the magnet? South pole towards? South pole away? Record your observations below.

**b.** Hold the magnet still near the loops of wire. Does the ammeter record a current? Record your observations.

**c.** Repeat part **a**. This time determine the direction of flow of positive charge through the solenoid based on the ammeter reading for each case:

- north pole moves toward the solenoid
- north pole moves away from the solenoid
- south pole moves toward the solenoid
- south pole moves away from the solenoid

For each case, determine if positive charge flows clockwise or counter-clockwise, when you look end on from the side the magnet is on. Record your data in the table below.

|  | Record if current flow is clockwise or counterclockwise from end on view |
|---|---|
| north pole moves toward the solenoid |  |
| north pole moves away from the solenoid |  |
| south pole moves toward the solenoid |  |
| south pole moves away from the solenoid |  |

Table 2. Table for Laboratory Exploration 1.2.

Ask your TA, if you are unsure of the direction positive charge flows through the circuit.

**d.** There is no battery in the circuit. Yet, charges are moving through the wire. There must be a force doing work on the charges. Further experiments show that the charges are moving because they are in an electric field. The changing magnetic field is creating an electric field. The work done per unit charge by the electric force is called the electromotive force (ξ) or emf. If more work is done per unit charge, the current is larger. The ξ and the current are related by ξ = IR.

The rate at which charges move through the wire seems to depend on the motion of the magnet. Further experiments reveal that the amount of current in the wire depends on:

- the time rate of change of magnetic flux
- the number of loops (turns) in the wire

The magnetic flux is defined the same way as the electric flux, but in terms of the magnetic field and the area, instead of the electric field and the area:

$$\Phi = BA\cos\theta$$

The equation for the ξ for a wire with N turns is:

$$E = -N\frac{\Delta\Phi}{\Delta t}$$

where N is the number of turns in the wire, $\frac{\Delta\Phi}{\Delta t}$ is the time rate of change of magnetic flux. This equation is called Faraday's Law.

The direction the current flows through the loop is such that the current in the loop will set up a magnetic field to oppose the change in flux through the loop. This is called Lenz's Law. Is this consistent with your data?

If the north pole of the magnet is moving toward the loop, the flux, which is essentially the number of field lines through the loop, is increasing. The current through the loop will set up a magnetic field in the opposite direction to the magnetic field produced by the north pole of the magnet in order to decrease the flux (the number of field lines) through the loop.

If the north pole of the magnet is moving away from the loop, the flux (number of field line through the loop) is decreasing. The current through the loop would set up a magnetic field in the same direction as the magnetic field produced by the north pole of the magnet in order to increase the flux (the number of field lines through the loop).

**e.** Use Lenz's Law to determine the direction that current flows through the loop when the south pole of the magnet moves toward the loop and when the south pole of the magnet is moved away from the loop.

Discuss your answers to part **e** with the TA.

**Equipment:**
- 1 oscillator
- 1 function generator
- 5 neodymium magnets
- 1 solenoid
- 1 Vernier Magnetic Field Sensor
- 1 Vernier Current probe
- 1 Vernier computer interface
- 1 Vernier computer software
- 1 multimeter

**Investigation 1: Using Faraday's Law to determine the magnetic field of a magnet**

**Investigation 1.1 Relationship between current and change in magnetic field strength**

In this Investigation, we will oscillate a magnet near a solenoid and observe the current through the solenoid. We will use the data to calculate the $\xi$ in two ways.

The setup consists of a function generator hooked to an oscillator. Five neodymium magnets are attached to the oscillator. The neodymium magnets will be placed just inside a solenoid placed next to the oscillator, as in the picture below.

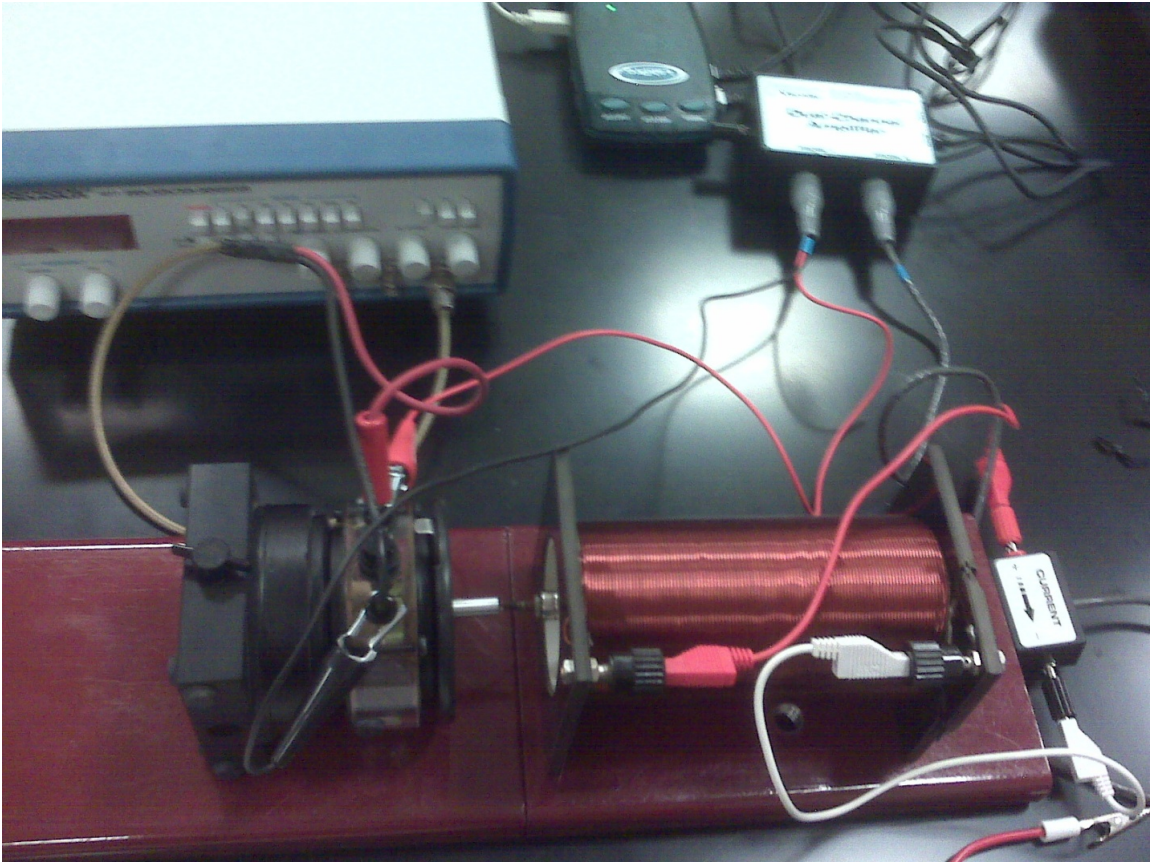

Figure 11: Picture of setup for Laboratory Investigation 1.1.

The function generator will be used to drive the oscillator back and forth. The driving wave function used will be a sine wave.

The oscillating magnets will cause a change in the magnetic field through the solenoid, changing the magnetic flux through the solenoid. We will use a magnetic Field Sensor to record the change in magnetic field as a function of time.

The changing magnetic field will produce an $\xi$. The $\xi$ will produce a current through the solenoid. We will measure the current through the solenoid using a current probe.

The $\xi$ can be calculated two ways:

(1) $\xi = IR$
(2) $\xi = -N \frac{\Delta \phi}{\Delta t}$

The current probe should be connected to the ends of the solenoid and to the computer interface. The Magnetic Field sensor should also be connected to the computer interface.

Make sure the magnetic field probe is set on the highest setting.

Open the computer interface program. On the Experiment drag down menu, set the sample time to 5seconds and the sample rate to 20 samples per second. Before taking data, you will need to zero the input. Use the zero button to zero both sensors.

Set the function generator to produce sine waves. Turn on the function generator and set the frequency to between 40-45 Hz.

Hold the magnetic field Sensor inside the solenoid and hit "Collect" to collect data.

Two graphs should appear on the screen – one of current vs. time and the other of magnetic field vs. time.

**a.** Write down Faraday's Law and use it to analyze the relationship between the graphs of magnetic field vs. time and current through the solenoid vs. time. How should they be related, based on Faraday's Law and how are they related graphically? Discuss.

**Discuss the relationship between the magnetic field vs. time and current through the solenoid vs. time with your TA before you continue.**

**b.** Measure the resistance of the solenoid wire. Record the resistance below.

**Investigation 1.2 Calculation of the ξ**

**a.** Determine the change in magnetic field at three points on the magnetic field curve. Choose a point at the top or bottom of the curves, a point of greatest slope and a third point. Use the software to determine the change in B as a function of time for each of the three points. Record the data in the table at the end of this section.

**b.** Calculate the ξ for each of the three points. Show your work below and record the results in the table at the end of this section.

**c.** Record the value of the current at each of the points in the table at the end of the section.

**d.** Calculate the ξ for each of the three points using ξ = IR. Record the results in the table below.

| Location | ΔB/Δt | $\xi = -N\dfrac{\Delta \varnothing}{\Delta t}$ | I | $\xi = IR$ |
|---|---|---|---|---|
| Top of curve | | | | |
| Point of greatest slope | | | | |
| Intermediate slope | | | | |

Table 3. Table for Laboratory Investigation 1.2.

Do the two calculations of the ξ agree or not? Explain.

**Equipment:**
        1 Induction wand
        1 Variable Gap Lab Magnet
        1 Large Rod stand
        145cm Long Steel Rod
        1 Multi Clamp
        1 Pasco Voltage Sensor
        1 Pasco Magnetic Field Sensor
        1 Pasco Rotary Motion Sensor
        1 mass Balance
        1 Meter Stick
        1 Science Workshop 500 Interface
        1 Data Studio Software

**Investigation 2: More work with Faraday's Law**

**Investigation 2.1**

In the setup at your table, a rigid pendulum with a coil at its end swings through a horseshoe magnet. An ξ is induced in the coil, as the magnet swings through the approximately constant magnetic field between the poles of the magnet. The setup is shown in the picture below.

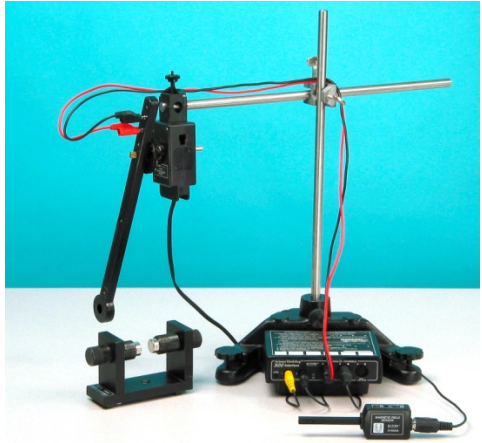

Figure12: Setup for Laboratory Investigation 2.1

**a.** Explain why there is an ξ in the coil as it passes through the magnetic field.

**b.** What do you expect a graph of ξ vs. time to look like? Sketch your prediction of a graph of ξ vs. time in the space below. Explain why you dew the graph the way you did.

**c.** Write out Faraday's Law in symbols in terms of the magnetic field, B, and the area, A. Is the field changing as the coil passes between the poles of the magnet? Is the area changing as the coil passes through the poles of the magnet? Explain.

**d.** What would you need to measure, in order to calculate the ξ as the coil passes between the poles of the magnet?

**e.** If you were in the frame of reference of the coil, is the field changing as the coil passes between the poles of the magnet? Is the area changing as the coil passes through the poles of the magnet? What would you need to measure, in order to calculate the ξ as the coil passes between the poles of the magnet?

**Investigation 2.2**

If the pole plates are not on the ends of the horseshoe magnet, as in the picture below, put them on now.

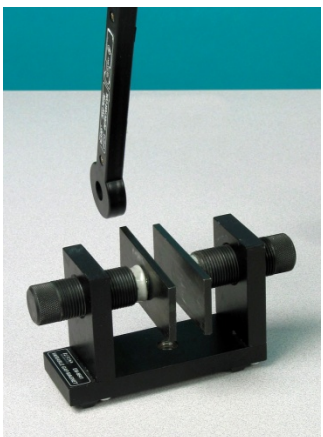

Figure13: Setup for Laboratory Investigation 2.2

Adjust the height of the coil so it is in the middle of the magnet. Align the wand from side-to-side so it will swing through the magnet without hitting it.

The Voltage Sensor should be plugged into Channel A of the ScienceWorkshop 500 interface. The Rotary Motion Sensor should be plugged into Channels 1 and the Magnetic Field Sensor should be plugged into Channel B.

**a.** Use the magnetic field Sensor to measure the strength of the magnetic field
between
the poles of the magnet. Record the value for the magnetic field strength in the space below.

Magnetic field strength________________

**b.** Determine the direction of the magnetic field between the poles of the magnet.
Which pole is the north pole? Describe how you determined the direction of
the magnetic field and record your results here.

**Investigation 2.3**

Plug the Voltage Sensor banana plugs into the banana jacks on the end of the coil
wand.
Drape the Voltage Sensor wires over the rods so the wires will not exert a torque on
the
coil as it swings. It helps to hold the wires up while recording data.

Open the DataStudio file called "Induced $\xi$".

**a.** Click START and pull the coil wand back and let it swing through the magnet.
Then
click STOP. The $\xi$ vs. time graph will show on the screen. You may use the Magnifier
Tool to enlarge the portion of the voltage vs. time graph where the coil passed
through

the magnet.

How does it compare to your prediction in **2.1.b**?

**b.** Discuss with your lab partners which part of the motion corresponds to which part of
the graph. Discuss why the ξ is positive, negative or zero at different points, as the coil
passes through the magnetic field. Record your thoughts here.

**Before you go on**, discuss with your TA which part of the motion corresponds to which
part of the graph.

### Investigation 2.4

The ξ is changing throughout the motion of the coil through the field. It is possible, however, to measure the average ξ for each part of the motion of the coil. We can also measure the average rate of change of the area of the coil or the average rate of change of the magnetic field for each part of the motion (if we look at Faraday's Law from the frame of reference of the coil).

**a.** Use the mouse to highlight the first peak on the graph and find the average ξ across
the coil. Record the average ξ and the relevant time interval in the space below.

average ξ____________        time interval______________

**b.** Calculate the average change in area during the relevant time interval in part **a.** Explain your calculation and show your work.

**c.** Calculate the average ξ from the average change in area. Show your calculation.

**d.** Compare the average ξ as measured in part **a.** to that calculated in part **c.**

### Investigation 2.5

It is also possible to calculate the average ξ from the frame of reference of the coil.

**a.** Calculate the average change in magnetic field during the relevant time interval in part
**2.4.a.** Explain your calculation and show your work.

**b.** Calculate the average ξ from the average change in area. Show your calculation.

**c.** Compare the average ξ as measured in part **2.4.a.** to that calculated in part **b.** How do they compare?

**Summary.** Summarize what you have learned about Faraday's Law and Lenz's Law in your own words. Include in your summary a discussion about how to determine the
direction of a current generated by a changing magnetic field, as well as a discussion of
different ways to create a changing magnetic flux.

# Laboratory Homework 9
# Magnetism III

1) Consider a U-shaped metal rod with a metal rod touching it, which is free to slide (shown below). A magnetic field is out of the page everywhere, as in the diagram below.

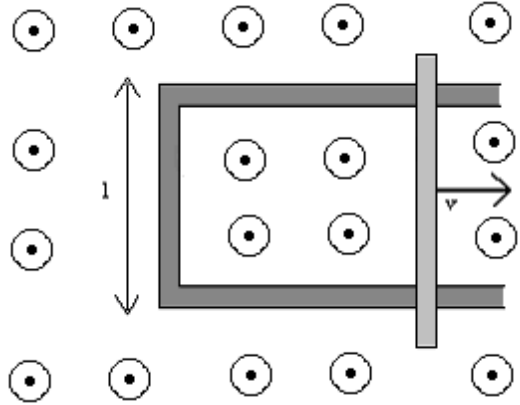

Figure 14: Picture for Laboratory Homework 9.

If the rod is moving to the right at a velocity v,

a) would there be a current? Explain. If so, determine the direction of the current.

b) would there be an ξ? Explain. If so, calculate the ξ in terms of the magnetic field, $B$, the velocity of the rod, $v$, and the length $l$. Show your work; do not just write down your result. Show your calculation.

# Appendix II

## A. Problem 1

### 1. Pre-test

1) Consider two parallel charged conducting plates, as in the diagram below. The top plate has a net negative charge and the bottom plate has a net positive charge of the same magnitude. The points A, B, C, and D represent points in space (not charges).

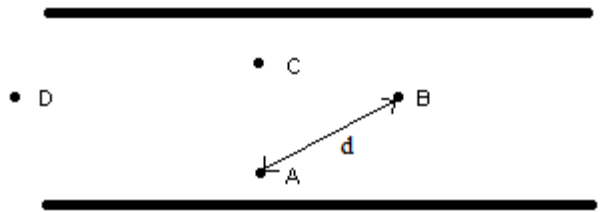

Figure 15: Picture for Problem 1 Pre-test.

   a) Rank the magnitude of the electric field at each point. Explain your ranking.
   b) How would you experimentally determine the electric field at point A? Explain.
   c) If the electric field at point A has a magnitude E, the potential difference between points A and B is $\Delta V$ and the distance between points A and B is d (as indicated in the diagram), which of the following statements is true? Circle your choice and explain your reasoning.
      i)     $E > \Delta V/d$
      ii)    $E = \Delta V/d$
      iii)   $E < \Delta V/d$

## 2. Post-test with a possible correct answer

1) Consider two parallel charged conducting plates, as in the diagram below. The top plate has a net negative charge and the bottom plate has a net positive charge of the same magnitude. The points A, B, C, and D represent points in space (not charges).

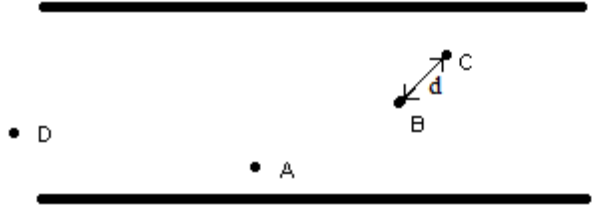

Figure 16: Picture for Problem 1 Post-test.

a) Rank the magnitude of the electric field at each point. Explain your ranking.

*For two large, charged conducting plates, the magnitude of the electric field is constant between the plates, except at the edges where the field is weaker, as there is a fringe field that bends outward and the field lines are less dense. The field lines point from the positive to the negative plate and are equally dense between the plates. The ranking is A = B = C > D.*

b) How would you experimentally determine the electric field at point B? Explain.

*Measure the potential difference between two points on either side of point B, in the direction of the electric field. The magnitude of the electric field can be found by dividing the potential difference between these two points by the distance between them. The field points directly upwards from the positive to the negative plate. Since the field is uniform between the plates, it is sufficient to measure the potential difference between any two equipotential lines and divide by the distance between them.*

c) If the electric field at point B has a magnitude E, the potential difference between points B and C is $\Delta V$ and the distance between points B and C is d (as indicated in the diagram), which of the following statements is true about the magnitude of the electric field? Circle your choice and explain your reasoning.
   i) $E > \Delta V/d$
   ii) $E = \Delta V/d$
   iii) $E < \Delta V/d$

*(i) $E > \Delta V/d$. The relevant distance to use in determining E is the distance between the equipotential lines through B and C, which is $\Delta x = d \cos \theta$, where $\theta$*

is the angle between the electric field and a vector **d** pointing from B to C. $E = \Delta V / d \cos \theta$, and since $d \cos \theta < d$, $E > \Delta V / d$.

## B. Problem 2

### 1. Part of a pre-test common among classes

1) Consider the circuit shown in the diagram (1a) below with two resistors in series. A third resistor is added to the circuit, as in diagram (1b).

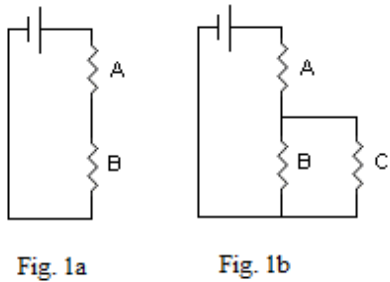

Fig. 1a      Fig. 1b

Figure 17: Picture for Problem 2 Pre-test.

a) Does the total resistance of the circuit increase, decrease or remain the same? Check the appropriate answer and explain your reasoning.

\_\_\_\_\_\_increases  \_\_\_\_\_\_decreases  \_\_\_\_\_\_remains the same

b) Does the current through resistor A increase, decrease, or remain the same after adding resistor C? Check the appropriate answer and explain your reasoning.

\_\_\_\_\_\_increases  \_\_\_\_\_\_decreases  \_\_\_\_\_\_remains the same

## 2. Part of a post-test common among classes with possible correct answers

1) Consider the circuit shown in the diagram (1a) below with two resistors in series. A third resistor is added to the circuit, as in diagram (1b).

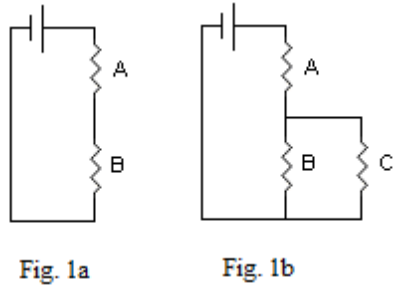

Fig. 1a          Fig. 1b

Figure 18: Picture for Problem 2 Post-test.

a) Does the current through the battery increase, decrease or remain the same? Check the appropriate answer and explain your reasoning.
   __X__ increase   ______ decrease   ______ remain the same

*When resistor C is added in parallel, the total resistance of the circuit decreases. This can be seen mathematically by solving for the resistance of the B-C network, $R_{BC}$, using $1/R_{BC} = 1/R_B + 1/R_C$. Then $R_{BC} = R_B R_C/(R_B + R_C)$ and $R_{BC} < R_B$. The total resistance decreases from $R_{Total} = R_A + R_B$ to $R'_{Total} = R_A + R_{BC}$. When the total resistance of the circuit decreases, the current through the circuit increases by Ohm's Law, $V = IR$, with V the battery voltage and R the total resistance of the circuit.*

b) Does the voltage across resistor A increase, decrease or remain the same? Check the appropriate answer and explain your reasoning.
   __X__ increase   ______ decrease   ______ remain the same

*Increases. The total current through the circuit passes through resistor A. Since that current increases, the current through A increases. Using Ohm's Law applied to resistor A, $V_A = I_A R_A$, with $R_A$ constant, an increase in $I_A$, increases the voltage across A, $V_A$.*

## C. Problem 3

### 1. Pre-test with possible correct answer

1) Which of the four graphs below is most likely a plot of the magnetic field vs. distance from a current-carrying wire? Base your choice on the shape of the graph, not the numerical values. Explain your choice.

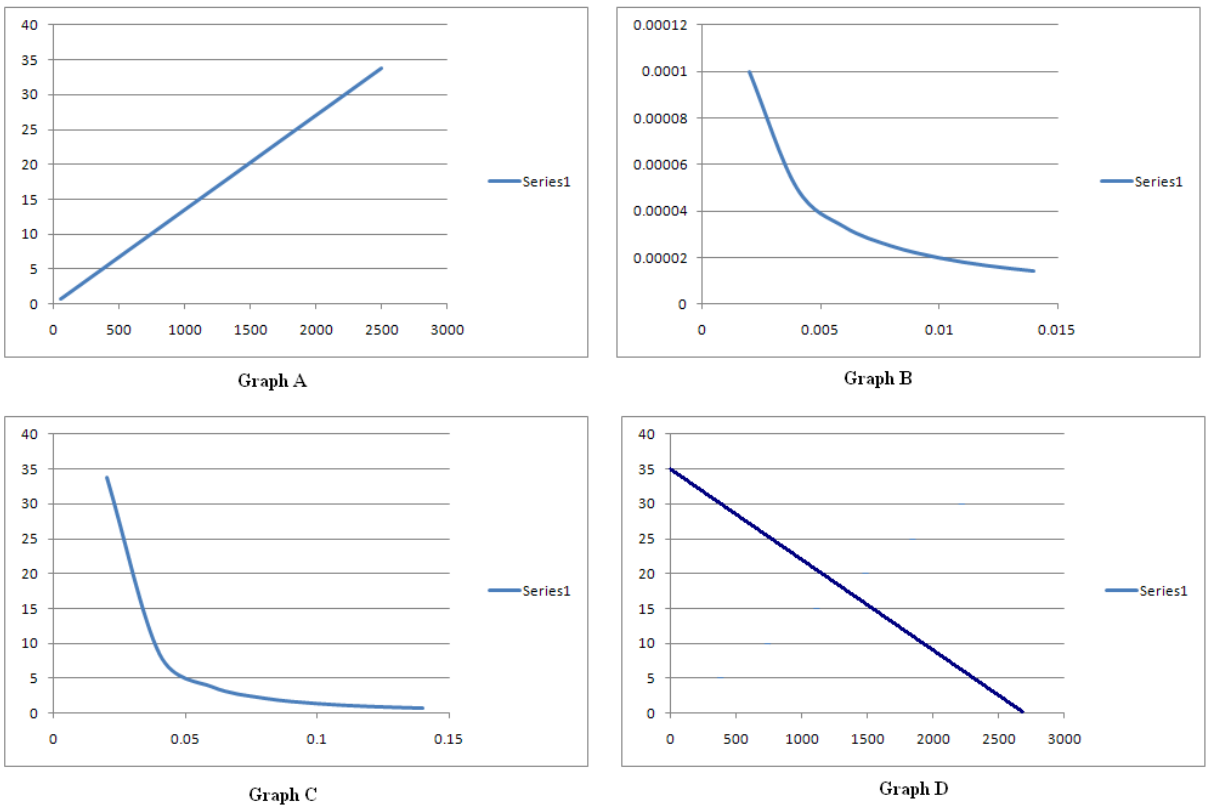

Figure 19: Graphs for Problem 3 Pre-test.

*Graph B. The magnetic field a distance r from a current carrying-wire drops off as 1/r. $B = \mu_0 I/2\pi r$. The decrease in the field with distance is not linear, eliminating Graph D, and it is a slower decrease than other common functions, such as $1/r^2$ or an exponential decrease. Graph B is more likely to represent a function of 1/r than Graph C, which drops off more quickly.*

## 2. Post-test with possible correct answer

1) The data in a laboratory got mixed up. There are two sets of data, shown below. One is the magnitude of the electric field a distance *r* from a point charge and the other is the magnitude of the magnetic field a distance *r* from a current-carrying wire, but the lab technicians don't know which set of data is which. What should they do to distinguish the two sets of data from each other and determine which is the electric field data and which is the magnetic field data? Describe in words what you would do to distinguish the two sets of data. You do not have to carry it out, just describe the process. The data are shown below.

| r | field | r | field |
|---|---|---|---|
| 0.02 | 33.75 | 0.002 | 0.0001 |
| 0.04 | 8.4375 | 0.004 | 0.00005 |
| 0.06 | 3.75 | 0.006 | 3.33E-05 |
| 0.08 | 2.109375 | 0.008 | 0.000025 |
| 0.1 | 1.35 | 0.01 | 0.00002 |
| 0.12 | 0.9375 | 0.012 | 1.67E-05 |
| 0.14 | 0.688776 | 0.014 | 1.43E-05 |

*Graph each set of data and use a graphing program to determine the best fit to the curves. The magnetic field should fit to 1/r and the electric field should fit t o 1/r².*

## D. Problem 4

### 1. Pre-test

1) Consider a loop of wire moving to the right with a constant velocity, v, through a region of magnetic field out of the page, as in the diagram below. The ammeter, A, measures the amount of current flowing through the wire.

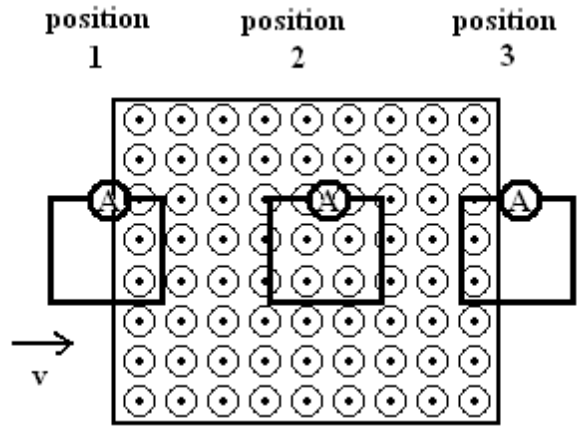

Figure 20: Picture for Problem 4 Pre-test.

a) In which case(s), position 1, 2 or 3, is there current flowing through the wire? Check off the case(s) in which current is flowing through the wire and *explain your reasoning*.

\_\_\_\_\_\_\_\_\_position 1     \_\_\_\_\_\_\_\_\_position 2     \_\_\_\_\_\_\_\_\_position 3

b) For each of the cases in part (a) in which you indicated that current was flowing through the wire, draw an arrow in the diagram to indicate the direction of the current. Explain your reasoning.
c) Rank any currents in order from greatest to least. Explain why you ranked them the way you did.

## 2. Post-test with possible correct answers

1) Consider three wire loops moving with a constant velocity, v, through a region of magnetic field out of the page, as in the diagram below. The wire loops in positions 1 and 2 are moving to the right. The wire loop in position 3 is moving upwards, towards the top of the page. An ammeter, A, in each loop measures the amount of current flowing through the wire.

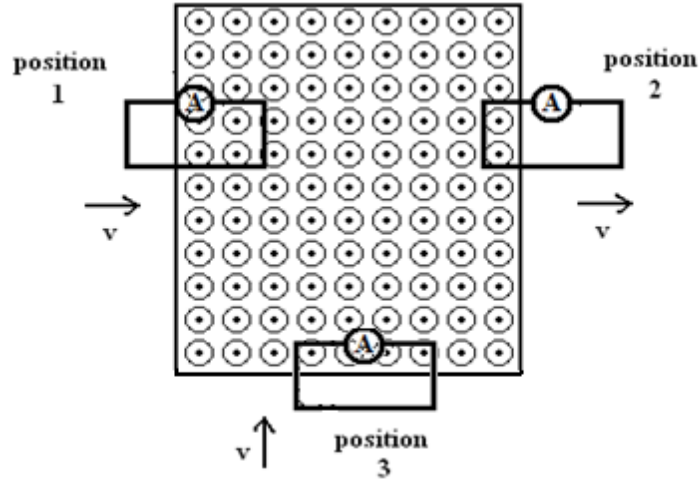

Figure 21: Picture for Problem 4 Post-test.

a) In which case(s), position 1, 2 or 3, is there current flowing through the wire? Check off the case(s) in which current is flowing through the wire and *explain your reasoning*.

____X____position 1     ____X____position 2     ___X_____position 3

*Current is flowing through all of the wires because the flux through each of the loops is changing with time. ξ = -NdΦ/dt.*

b) For each of the cases in part (a) in which you indicated that current was flowing through the wire, draw an arrow in the diagram to indicate the direction of the current. Explain your reasoning.

*Clockwise in positions 1 and 3 and counterclockwise in position 2. The induced current flows in the direction that will create an induced magnetic field and an induced flux that opposes the change in flux due to the original field and the moving coils.*

c) Rank any currents in order from greatest to least. Explain why you ranked them the way you did.

*Assume identical loops. Then ξ=IR and the current is proportional to the ξ. The rate of change of flux is the same in positions 1 and 2, so they will have the same magnitude ξ, since ξ= -NdΦ/dt. In loop3, flux changes at a greater rate, since more field lines enter the loop per unit time. So 3 > 1 = 2.*

# Appendix III
# Research Rubrics

## A. Problem 1

### 1. Part a.

**Completely correct.** To be completely correct, students had to rank the points as A=B=C>D, recognize that the electric field is uniform inside the plates and that the field lines are less dense outside the plates. D, lying outside the plates has a weaker field than the other three points.

**Partially correct.** An answer was counted partially correct, if it was stated that the field was constant between two large charged, conducting plates, and therefore the electric field at all of the points were equal.

### 2. Part b.

**Completely Correct.** To be completely correct, the answer needed to describe the use of a voltmeter and the placement of the probes to measure a potential difference in the direction of the electric field and then identify the correct distance to be measured and used in the equation $E = -\Delta V/\Delta x$.

Note: There were students who described various experiments using charged balls or particles (usually not based on available introductory lab equipment). However, none of them were completely correct.

**Partially correct.** Partially correct answers described a voltmeter measurement between two points, but did not explicitly explain that the measurement should be made in the direction of the electric field. For example, a student would describe placing one of the voltmeter leads on one of the plates and the other at point B, recording the voltage and dividing by the distance between the plate and the point, without explicitly stating that the distance measurement should be made parallel to the electric field lines.

### 3. Part c.

**Completely correct.** A completely correct answer chose (i) $E > \Delta V/d$ and correctly described that the magnitude of E is $E = \Delta V/\Delta x$, where $\Delta x$ is the distance between equipotential lines and $\Delta x = d \cos \theta$, where $\theta$ is the angle between the electric field and a vector **d** pointing from B to C.

**Partially correct.** Partially correct answers were usually incomplete answers. For example, the student would state the correct distance needed, but not explain why that would make $E > \Delta V/d$. They assumed that stating the correct distance needed was a sufficient explanation.

## B. Problem 2

### 1. Part a.

**Completely correct.** To be completely correct a student had to demonstrate qualitatively or through calculations that the resistance of the B-C network is less than the resistance of B and that the total resistance decreases. Then they had to correctly apply Ohm's Law, V = IR, with V the battery voltage and R the total resistance of the circuit.

**Partially correct.** A student was awarded partial credit for an answer with the correct process, but an algebraic error or an explanation with a minor error in wording or one that lacked sufficient detail.

### 2. Part b.

**Completely correct.** To be completely correct a student had to correctly apply Ohm's Law or a combination of Ohm's Law and one of Kirchhoff's Laws (there are a number of correct ways to work the problem). One of the simplest ways to work the problem is to recognize that the current through A increases and demonstrate qualitatively or mathematically through a correct application of Ohm's law applied to resistor A, that the voltage across A increases because the current through resistor A increases, but the resistance remains the same.

**Partial Correct.** Answers were counted as partially correct for the use of Ohm's law with only minor calculation errors or explanations with errors in wording or that lacked sufficient detail.

## C. Problem 3

### 1. Pre-test

**Completely correct.** To be completely correct, the student must choose the correct graph and make a correct argument for why that is the correct graph. The argument should include the recognition that $B \propto 1/r$ and how this is represented by their choice of graph.

**Partially correct.** A choice of Graph B supported with an explanation of the features of the graph, such as a slow decrease, without an explicit discussion of the difference between Graph B and Graph C was counted as partially correct.

### 2. Post-test

**Completely correct.** A discussion of graphing the data, finding the best fit to each curve, and choosing the magnetic field as the best fit to $1/r$ and the electric field as the best fit to 1/r² or a discussion of how to substitute the data into the known equations for magnetic and electric fields to determine which set of data fits which equation. There are other possible ways to work the problem, such as taking the ratio of the field values at different radii to determine if the field values are decreasing by r or r².

**Partially correct.** There were not many partially correct answers. Answers that lacked sufficient detail were counted as partially correct.

### D. Problem 4

#### 1. Part a

**Completely correct.** To have a completely correct answer, the student must recognize that the flux through all of the loops is changing and that it is the changing flux that is inducing a current in the loops.

**Partially correct.** There were incomplete arguments that addressed the area within the magnetic field changing with time, without explicitly mentioning flux, that were considered partially correct.

#### 2. Part b

**Completely correct.** For an answer to be considered completely correct, the student had to demonstrate an understanding of the induced current creating an induced magnetic field that generated an induced flux that opposed the change in flux for all three loops.

**Partially correct.** There were incomplete arguments of the induced field opposing the magnetic field, that were counted as partially correct.

#### 3. Part c

**Completely correct.** In order to be considered a completely correct answer, the student had to recognize that the flux through loop 3 was changing at a faster rate than the flux through the other two loops and that the flux through loops 1 and 2 was changing at the same rate. Correct answers could be expressed in words or equations. $\Delta\Phi = \Delta(BA)/\Delta t$, where $A = xy$. For loops 1 and 2, $y$ is constant and $v = \Delta x/\Delta t$, so $\Delta\Phi = Byv$. For loop 3, $x$ is constant and $y$ is changing, resulting in $\Delta\Phi = Bxv$.

**Partially correct.** Most students wrote equations with little or no explanation; as a result there were very few partially correct answers.

# Appendix IV
# Common Student Incorrect Responses

## A. Problem 1

We present here some of the common student incorrect responses.

1. **Part a**

    The most common error in *part a* was to rank the electric field magnitude by the proximity to one of the plates. Students either ranked the field magnitude as higher closer to the positive plate or higher closer to the negative plate. Over 70% of the incorrect answers fell into this category. A handful of students ranked the field higher in the middle, decreasing in magnitude as you moved closer to either of the plates.

2. **Part b**

    *Part b* incorrect answers were characterized by vagueness, lack of detail and incompleteness. The students did not have the ability to identify the key measurements they had made in the previous week's lab and write a clear and concise description of the measurements and how to calculate the magnitude of the electric field. Examples of vague and incomplete answers follow. (These are complete answers. Not parts of answers.)

    *"Place a voltmeter at point B and on the positive plate."*

    *"By determining the necessary variables and solving for E using the proper equation."*

    *"You would get d, the distance between them and then get the potential difference between the two points and divide."*

    *"I would use a voltmeter to measure the voltage at different points. Then after obtaining the distance d, I would solve for the magnitude."*

3. **Part c**

    The most common incorrect answer in *part c* was to choose *(ii)* and argue that $E = \Delta V/d$ was the equation learned in class. Typical explanations were:

    *"The answer is ii. In class and in the homework, we learned that $E = \Delta V/d$."*

    *"This ($E=\Delta V/d$) is the defined operation."*

## B. Problem 2

1. **Part a**

    The most common incorrect answer was to argue that the total resistance increases when a resistor is added and that the current through the battery therefore decreases. Some examples are:

    *"Current decreases because there is more resistance."*

    *"The current will get divided and there is more resistance, so it decreases."*

    *"The resistors slow the flow of the circuit. The electrical potential is lost with the resistance of a new resistor."*

    The next most common argument was that the current remains constant, independent of the number of resistors. Some examples are:

    *"The current doesn't change and it is wired in parallel."*

    *"The current through a given circuit is constant throughout with a given battery."*

    *"Current is constant."*

2. **Part b**

    The most common incorrect answer in part b was that the voltage remains the same because it is independent of the resistors connected.

    Examples of this argument are:

    *"Voltage is constant and unaffected, it is current that is affected by resistors. The battery produces a voltage that is independent of resistors in parallel."*

    *"The voltage is unaffected by resistors."*

C. **Problem 3**

1. **Pre-test**

    The most common incorrect explanations for the pre-test were to explain choices of graphs B or C as being correct because they described the exponential decay of the magnetic field. Examples of this are:

    *"It drops exponentially, but I'm not sure if B or C."*

*"Looks most like exponential."*

*"Field strength is exponentially related to distance."*

The students who chose the correct answer usually were not able to correctly explain it and usually used exponential decay arguments. The most common CI error was to choose graph B and describe it as an exponential decrease that is not as fast a decrease as graph C. An example of this kind of answer is:

*"Graph B fits a more true exponential graph of magnetic field (B) vs. distance since magnetic field decreases as distance increases. It also has a less steeper slope than graph C."*

The second most common error, after exponential decay, was to argue a linear decrease. Examples of this argument are:

*"Graph D because as distance increases, magnetic field decreases proportionally."*

*"Because the relationship of the magnetic field to distance is linearly."*

Many students wrote down the correct relationship or correct formula for the magnetic field and described the function as linear. Examples of this are:

*"Graph D: B = 1/R The relationship is linear and the further away you get, the smaller the value."*

*"Graph D. B = $\mu_0 I/2\pi r$."*

2. **Post-test**

   One of the most common errors was a lack of detail. In particular, students would argue that you simply plug the numbers into the formula and see which data fit. This does not address the fact that you don't know the current, I, or the charge, Q, and you need to take a ratio or determine the constant in order to determine how the field data decreases with distance. Examples of this are:

   *"To distinguish the data, you could plug both into the formula B = $\mu_0 I/2\pi r$ and see which sets of data fit better, this would establish which is the electric and magnetic field."*

   *"I would use the r values into $\mu_0 I/2\pi r$ to see which is which, that way you could distinguish them."*

Another common incorrect answer was to interpret the question as asking the student to determine which set of data is from measurements of an electric field and which is from measurements of a magnetic field, not the process of how you would determine it. Students made arguments, even though no units are given, that the either the electric field or the magnetic field would produce larger values. Examples of this type of argument are:

*"The electric field data should have much lower values compared to the magnetic field data when same distance of radius is used. Therefore the field on the right is the electric field and left shows magnetic field."*

*"The magnitude of the electric fields are gonna be a larger number than the magnetic. As the further the distance the smaller the fields."*

*"The smaller field values are those of the magnetic field?"*

Another common answer was that you should re-do the experiment.

**D. Problem 4**

**1. Part a**

Many of the incorrect answers reflected a conception that the loop simply had to be in, or partially in, the field but not that the flux had to be changing for there to be current.

*"They are all in the magnetic field."*

*"All loops have a portion of their area in the magnetic field, so there is no reason current will flow through one of them and not the other."*

Some students thought the location of the ammeter determined whether current flowed, i.e., if the ammeter was inside the field, there would be current.

*"The ammeter, along with the positioning, allows the current to flow in the loop as they will have induced current by the magnetic field."*

*"Since there is an ammeter, there is current flowing through the wire, having the magnetic field going through the coil to give current."*

**2. Part b**

Many students used the right hand rule incorrectly Examples of this are:

*"With regards to the right hand rule, since the magnetic field is coming out of the page, the current flows to the right."*

*"Clockwise, I attempted to use the right hand rule."*

*"Using the right hand rule, where our thumb is the current, the mag field points into the page so the current is clockwise."*

*"If the velocity is going to the right, then according to the right hand rule, the current is directing upwards in the same direction as the force."*

Some students did not understand that the external magnetic field was not produced by the currents in the loops.

*"There is a current going counterclockwise which creates magnetic field lines coming out of the paper which counteracts the increase in the magnetic field's magnitude."*

3. **Part c**

   The answers varied widely; there were not common incorrect answers.